\documentclass[twocolumn,aps,pre]{revtex4}
\usepackage{graphicx,graphics,color,epsfig}
\usepackage{amsmath}
\usepackage{amssymb}

\begin{document}
\title{
A strategy for solving difficulties in spin-glass simulations
}
\author{Tota Nakamura}
\affiliation{%
Faculty of Engineering
Shibaura Institute of Technology, %
307 Fukasaku, Minuma, Saitama 337-8570, Japan}

\date{\today}

\begin{abstract}
A spin-glass transition has been investigated for a long time
but we have not reached a conclusion yet
due to difficulties in the simulation studies.
They are slow dynamics, strong finite-size effects, 
and sample-to-sample dependences.
We found that a size of the spin-glass order
reaches a lattice boundary within a very short Monte Carlo step.
A competition between the spin-glass order and a boundary condition
causes these difficulties.
Once the boundary effect was removed, physical quantities
exhibited quite normal behaviors.
They became self-averaging in a limit of large replica numbers.
A dynamic scaling analysis on nonequilibrium relaxation functions 
gave a result that
the spin-glass transition and the chiral-glass transition occurs
at the same temperature in the Heisenberg model in three dimensions.
The estimated critical exponent $\nu$ agrees with the experimental result.
\end{abstract}

\maketitle

\section{Introduction}
\label{sec:intro}

A spin glass(SG)\cite{SGReview1,SGReview2,SGReview3,KawashimaRieger,Kawamurareview}
is a random magnet consisting of
ferromagnetic interactions and antiferromagnetic interactions 
distributed randomly.
It shares many common interest and difficulties with other random systems.
A glassy state appears at low temperatures.
A motion of each spin is particularly slow, and there is no spatial order.
This situation produces nontrivial and exotic magnetic states and
has been attracting much interest.
A spin-glass study has been a challenging field
of developing an efficient numerical algorithm.
One successful achievement is
the temperature-exchange method.\cite{hukutempexchange}
It is now a standard algorithm in SG simulations and 
applied to various complex systems.
A quantum-annealing algorithm\cite{KadowakiNishimori} was developed 
to obtain the ground state of the SG system.
It is now considered as a practical solution for various 
non-convex optimization problems.

Although the algorithms developed and theoretical investigations
for more than 30 years, 
there still remain many arguments unsettled in the SG study.
This is because there are difficulties in simulation studies.
Namely,
the simulations suffer from severe slow dynamics, and it takes a very 
long time to equilibrate the system.
We also need to take averages of physical quantities over different 
realizations of random bond configurations.
A sufficient sample number increases when there are strong
sample-to-sample dependences.
Then, a more computational time is needed, and we can simulate
only small-lattice systems.
The obtained data include strong finite-size effects, and
a finite-size scaling analysis encounters large
finite-size corrections.
A final conclusion sometimes depends on the way how we treat 
the correction terms.
These are common difficulties in random systems.

In this paper, we focus on a problem whether
the SG transition in the Heisenberg model is driven by the
spin degrees of freedom or the chirality degrees of freedom.
The Heisenberg SG model is the first approximation for
the canonical SG materials.
An origin of the debate on this model
dates back to a work by Olive, Young, and Sherrington\cite{Olive} in 1986,
where the SG transition was not observed by the Monte Carlo(MC) simulations.
The simulations were performed up to a linear lattice size $L=32$.
In 1992, Kawamura \cite{KawamuraH1,KawamuraReview} 
introduced the chirality scenario, wherein
the SG transition observed in real materials was considered as
an outcome of the chiral-glass(CG) transition without the SG transition.
A finite spin anisotropy was considered to mix the SG order and the CG order. 
Its counterargument is an existence of a simultaneous SG and CG 
transition, which was observed by MC simulations after 2000's.%
\cite{matsubara1,matsubara2,matsubara3,nakamura,Lee,berthier-Y,%
picco,campos,Lee2,shirakuraB,totawindow}
However, the results supporting the chirality scenario were also
reported at the same time.%
\cite{matsumoto-huku-taka,HukushimaH,HukushimaH2}
In 2009, two studies\cite{fernandez, viet,viet2}
in both sides drew two opposite conclusions even though the authors 
performed similar amounts of simulations,
but treated the finite-size effects differently.
The linear sizes were
$L=8-48$\cite{fernandez} and $L=6-32$\cite{viet,viet2}.
This disagreement suggests that
a strong finite-size effect hopelessly prevented us to
reach a conclusion of this problem.

Previous simulation studies mostly
applied the equilibrium MC method
and the finite-size scaling analysis.
They also imposed the periodic boundary conditions.
This strategy was first developed in uniform spin systems.
However, there is no translational symmetry in spin glasses.
Periodic boundary conditions are incompatible with the SG order.
This incompatibility produces strong sample dependences and 
strong finite-size effects.
In order to solve the difficulties in SG simulations, 
we need to reexamine this strategy.
In this paper, we clarify an origin of the difficulties, 
and propose a strategy to solve them.

The present paper is organized as follows.
Section \ref{sec:model} describes the model we treat in this paper.
We also give expressions for observed physical quantities.
In Sec. \ref{sec:diff}, 
we clarify an origin of the simulation difficulties.
In Sec. \ref{sec:method}, we introduce our strategy.
In Sec. \ref{sec:results}, numerical results are presented.
Section \ref{sec:discussion} is devoted to summary and discussions.

\section{Model and observables}
\label{sec:model}

A Hamiltonian of the present model is written as follows.
\begin{equation}
  {\cal H} = - \sum_{\langle ij \rangle} 
     J_{ij} \mbox{\boldmath $S$}_i \cdot 
            \mbox{\boldmath $S$}_j 
\label{eq:H1}
\end{equation}
The sum runs over all the nearest-neighbor spin pairs $\langle ij \rangle$.
The interactions, $J_{ij}$, take Gaussian variables with a zero mean and
a standard deviation, $J$.
The temperature, $T$, is scaled by $J$.
Linear lattice size is denoted by $L$. 
A total number of spins is $N = L \times (L-1)^2$, and
skewed periodic boundary conditions are imposed.

We calculated in our simulations
the SG and the CG susceptibility, 
$\chi_\mathrm{SG}$ and $\chi_\mathrm{CG}$, 
and
the SG and the CG correlation functions,
$f_\mathrm{SG}$ and $f_\mathrm{CG}$,
from which we estimated the SG and the CG correlation lengths, 
$\xi_\mathrm{SG}$ and $\xi_\mathrm{CG}$.
We evaluated these quantities at MC steps, $t$, with a same interval
in a logarithmic scale, namely at $t=10^{0.05i}$ with an integer $i$.

The SG susceptibility is defined by the following expression.
\begin{equation}
\chi_\mathrm{SG}\equiv\frac{1}{N}
\left[
\sum_{i,j}\langle \mbox{\boldmath $S$}_i \cdot \mbox{\boldmath $S$}_j \rangle^2
\right]_\mathrm{c}
\label{equ:xsg}
\end{equation}
The thermal average is denoted by $\langle \cdots \rangle$, and 
the random-bond configurational average is denoted by $[ \cdots ]_\mathrm{c}$.
The thermal average is replaced by an average over independent real
replicas that consist of different thermal ensembles:  
\begin{equation}
\langle \mbox{\boldmath $S$}_i \cdot \mbox{\boldmath $S$}_j \rangle
= \frac{1}{m}\sum_{A=1}^m 
\mbox{\boldmath $S$}_i^{(A)} \cdot \mbox{\boldmath $S$}_j^{(A)}.
\label{eq:therave}
\end  {equation}
The superscript $A$ is a replica index.
A replica number is denoted by $m$.
We prepare $m$ real replicas for each random-bond configuration
with a different initial spin state.
Each replica is updated using a different random number sequence.
A replica number controls an accuracy of the thermal average.

An overlap between two replicas, $A$ and $B$, is defined by
\begin{equation}
  q_{\mu \nu}^{A B} \equiv \frac{1}{N}
   \sum _i {S}_{i\mu}^{(A)} {S}_{i\nu}^{(B)}.
\label{equ:qsg}
\end{equation}
Here, subscripts $\mu$ and $\nu$ represent three components of Heisenberg
spins: $x, y$, and $z$.
The SG susceptibility is rewritten using this overlap as
\begin{equation}
\chi_\mathrm{SG} = \frac{N}{C_m}
\left[
\sum_{A>B, \mu,\nu}
(q^{AB}_{\mu\nu})^2
\right]_\mathrm{c}.
\end  {equation}
Here, $C_m=m(m-1)/2$ is a combination number of choosing two replicas
out of $m$ replicas.
Similarly, the CG susceptibility is defined by 
\begin{equation}
  \chi_\mathrm{CG} \equiv \frac{3N}{C_m}
\left[
\sum_{A > B}
(q_{\kappa}^{A B})^2
\right]_\mathrm{c} ,
\label{equ:xcg}
\end{equation}
where
\begin{eqnarray}
  q_{\kappa}^{A B}&\equiv&\frac{1}{3N}\sum _{i,\mu} \,
     {\kappa}_{i,\mu}^{(A)} {\kappa}_{i,\mu}^{(B)},
\label{equ:qcg}
\\
 \kappa _{i,\mu}^{(A)}&\equiv &
\mbox{\boldmath $S$}_{i+\hat{\mbox{\boldmath $e$}}_{\mu}}^{(A)}
\cdot
(
\mbox{\boldmath $S$}_{i}^{(A)}
\times
\mbox{\boldmath $S$}_{i-\hat{\mbox{\boldmath $e$}}_{\mu}}^{(A)}).
\label{equ:localk}
\end{eqnarray}
This $\kappa _{i,\mu}^{(A)}$ is a local scalar chirality,
where $\hat{\mbox{\boldmath $e$}}_{\mu}$ denotes a unit lattice vector along
the $\mu$ axis.

An SG correlation function is defined by the following expressions.
\begin{eqnarray}
f_\mathrm{SG}(r) &\equiv& \left[
\frac{1}{N}
\sum_i^N\langle
 \mbox{\boldmath $S$}_i \cdot \mbox{\boldmath $S$}_{i+r} \rangle^2\right]_\mathrm{c}
\label{eq:sgcordef}
\\
&=& \left[ 
\frac{1}{NC_m}
\sum_{A>B, i, \mu,\nu}
q_{\mu\nu}^{AB}(i)
q_{\mu\nu}^{AB}(i+r)
\right]_\mathrm{c}
\label{eq:sgcorover}
\\
&=&\left[
\frac{1}{N}
\sum_i^N
\left(
\frac{1}{m} 
\sum_{A=1}^m
\mbox{\boldmath $S$}_i^{(A)} \cdot \mbox{\boldmath $S$}_{i+r}^{(A)}
\right)^2
\right]_\mathrm{c}
\label{eq:sgcoruse}
\end{eqnarray}
When a replica number is two, it is equivalent to 
the four-point correlation function as shown in Eq.~(\ref{eq:sgcorover}).
Since we will use a large replica number up to 72 in this study, 
it is very time-consuming to take an 
average over $C_m$ different overlap functions.
Therefore, we took another expression (\ref{eq:sgcoruse}).
For a given distance $r$ and a site $i$, we calculated a spin correlation 
function for each replica $A$, and store it in an array memory.
Then, a replica average is taken and the value is squared.
We obtain the correlation function $f_\mathrm{SG}(r)$
by taking an average of the squared value over lattice sites $i$.
Changing a value of $r$ with the same procedure, we finally 
evaluated all the correlation functions.
A total calculation time is reduced by this procedure because
the maximum value of $r=L/2-2$ is much smaller than $C_m$.
Here, we considered the correlations for three directions, 
$(1,0,0)$, $(0,1,0)$, and $(0,0,1)$, 
and took an average over them.
We obtained a CG correlation function in a same manner
replacing the local spin variables with the local chirality variables:
\begin{eqnarray}
f_\mathrm{CG}(r) 
&=&\left[
\frac{1}{3N}
\sum_{i,\mu}^N
\left(
\frac{1}{m} 
\sum_A^m
\kappa_{i,\mu}^{(A)} \kappa_{i+r,\mu}^{(A)}
\right)^2
\right]_\mathrm{c}.
\label{eq:cgcoruse}
\end{eqnarray}
A unit of three neighboring spins in a same direction is considered 
and values for three directions are averaged.

\begin{figure}
  \resizebox{0.40\textwidth}{!}{\includegraphics{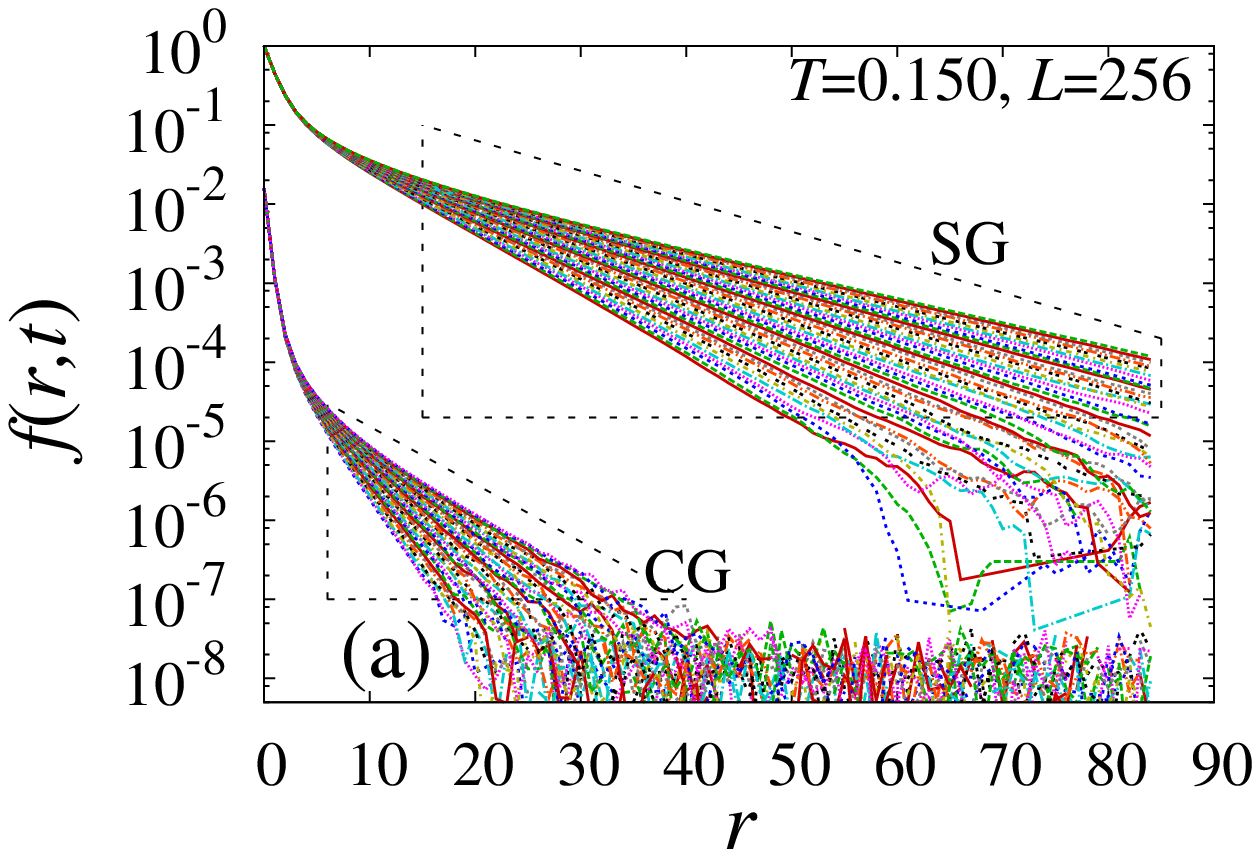}}
  \resizebox{0.23\textwidth}{!}{\includegraphics{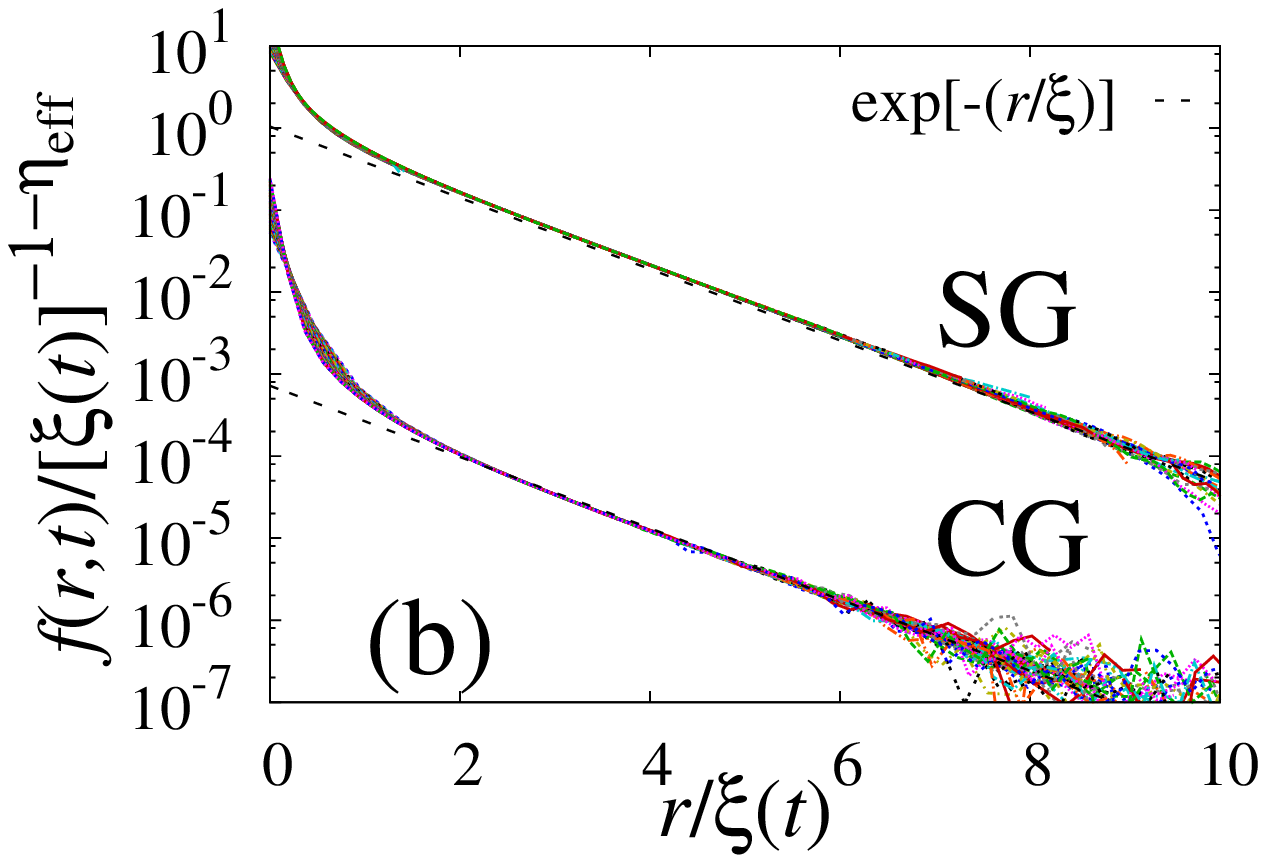}}
  \resizebox{0.23\textwidth}{!}{\includegraphics{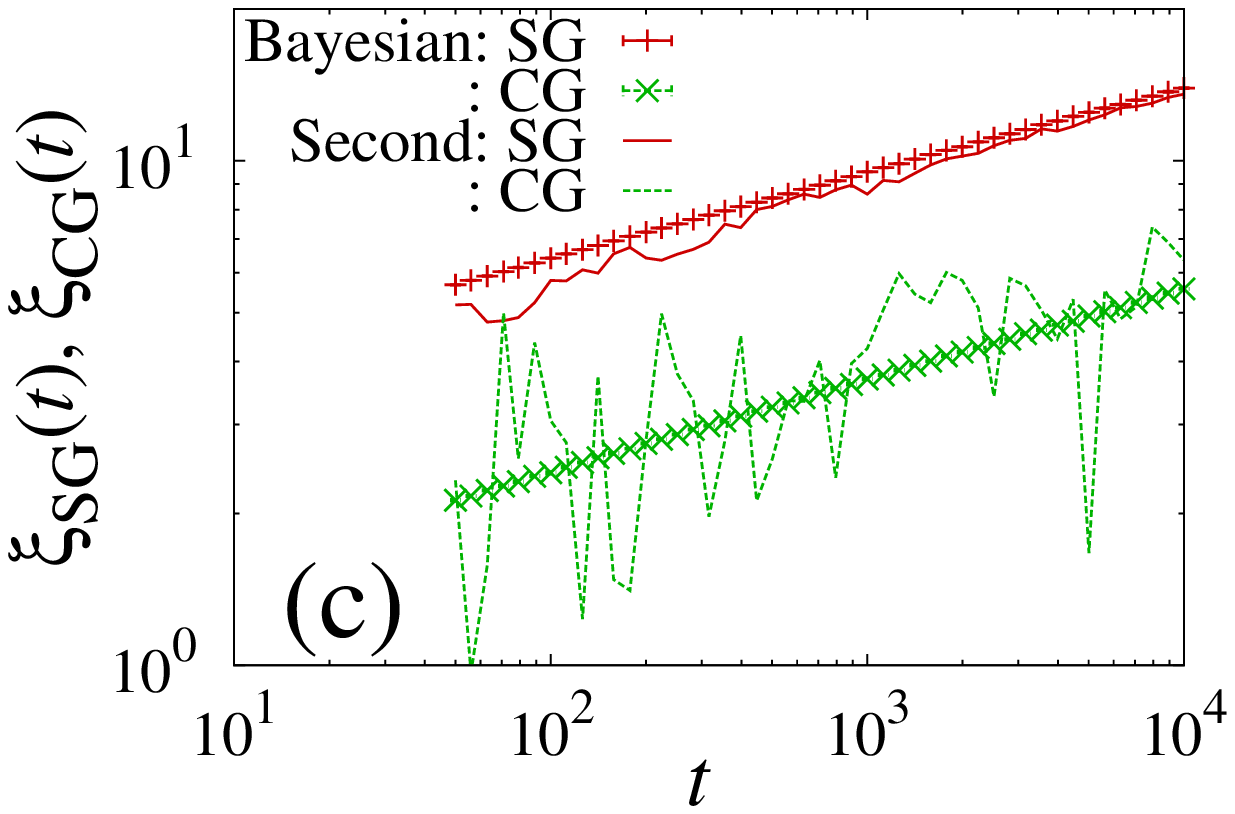}}
  \caption{
(Color online)
(a)
SG- and CG-correlation functions at time steps ranging from 50 to 10000
when $T=0.15$ and $L=256$.
Data within the dotted lines are used to estimate the correlation length.
(b)
A scaling plot of the correlation functions.
Effective exponents are 
$\eta_{\rm eff}=0.221$ for SG and $\eta_{\rm eff}=0.593$ for CG.
(c)
Symbols with error bars depict estimated data of the correlation length.
We also plotted with lines results obtained by the second-moment method.
}
\label{fig:fscale}
\end{figure}

In most simulation studies,
a correlation length has been estimated using the second-moment 
method, $\xi=\sqrt{\chi_0/\chi_k-1}/k$,\cite{cooper}
where $\chi_0$ denotes the susceptibility and 
$\chi_k$ denotes the Fourier transform
of the susceptibility with the smallest wave number, $k$.
As a system size increases, $\chi_k$  approaches $\chi_0$ and $k$
approaches zero.
Then, an estimated value of $\xi$ includes a large statistical error
by a situation of $0/0$.
On the other hand, the value includes a systematic error, which is
on the order of $\xi/L$, when a lattice size is small. 
In this paper, we estimated the correlation 
length by the Bayesian inference\cite{totabayes} using the data
of correlation functions.
The Bayesian theorem exchanges a prior probability and 
a posterior probability.
For example, let us suppose that a correct correlation length, 
$\xi(t)$, was obtained at each MC step, $t$.
Because of the critical scaling hypothesis,
the correlation function $f(r)$ behaves as $r^{-d+2-\eta}$,
where, $d$ is a dimension and $d=3$ here.
If we scale $r$ by the correlation length $\xi(t)$, 
a correlation function at each step, $f(r, t)$, 
is rescaled by $\xi^{-1-\eta}(t)$.
Therefore, the correlation function data should be scaled by 
plotting $f(r,t)/\xi^{-1-\eta}(t)$ versus $r/\xi(t)$.
Now, we use the Bayesian theorem and exchange the argument.
Proper $\xi(t)$ and $\eta_{\rm eff}$ can be obtained as scaling parameters
such that the scaling plot became the best.
This inference procedure is performed 
by the kernel method.\cite{kernel,harada}
An effective exponent $\eta_{\rm eff}$ depends weakly on the temperature
reflecting the corrections to scaling.
It is expected to coincide with the critical exponent if
the temperature is the critical temperature.

Figure~\ref{fig:fscale}(a) shows data of the correlation functions.
In an inference procedure, we discarded data of short MC steps($t<50$),
data of short-range correlation ($r<15$ for SG, and $r<6$ for CG),
data near the boundary ($r>L/3$),
and data of small $f(r, t)$ values
($f(r, t)<2\times 10^{-5}$ for SG and $f(r, t)<1\times 10^{-7}$ for CG).
A result of the scaling is shown in Fig.~\ref{fig:fscale}(b).
All the data ride on a single line.
The estimated correlation-length data are plotted with symbols in
Fig.~\ref{fig:fscale}(c).
Error bars are negligible.
We also plotted with lines results obtained by the second-moment method. 
The data fluctuate much and we cannot study the 
behavior of relaxation functions with them.

\section{Difficulties in spin-glass simulations}
\label{sec:diff}

\subsection{Finite-size effects}

\begin{figure}
  \begin{center}
  \resizebox{0.40\textwidth}{!}{\includegraphics{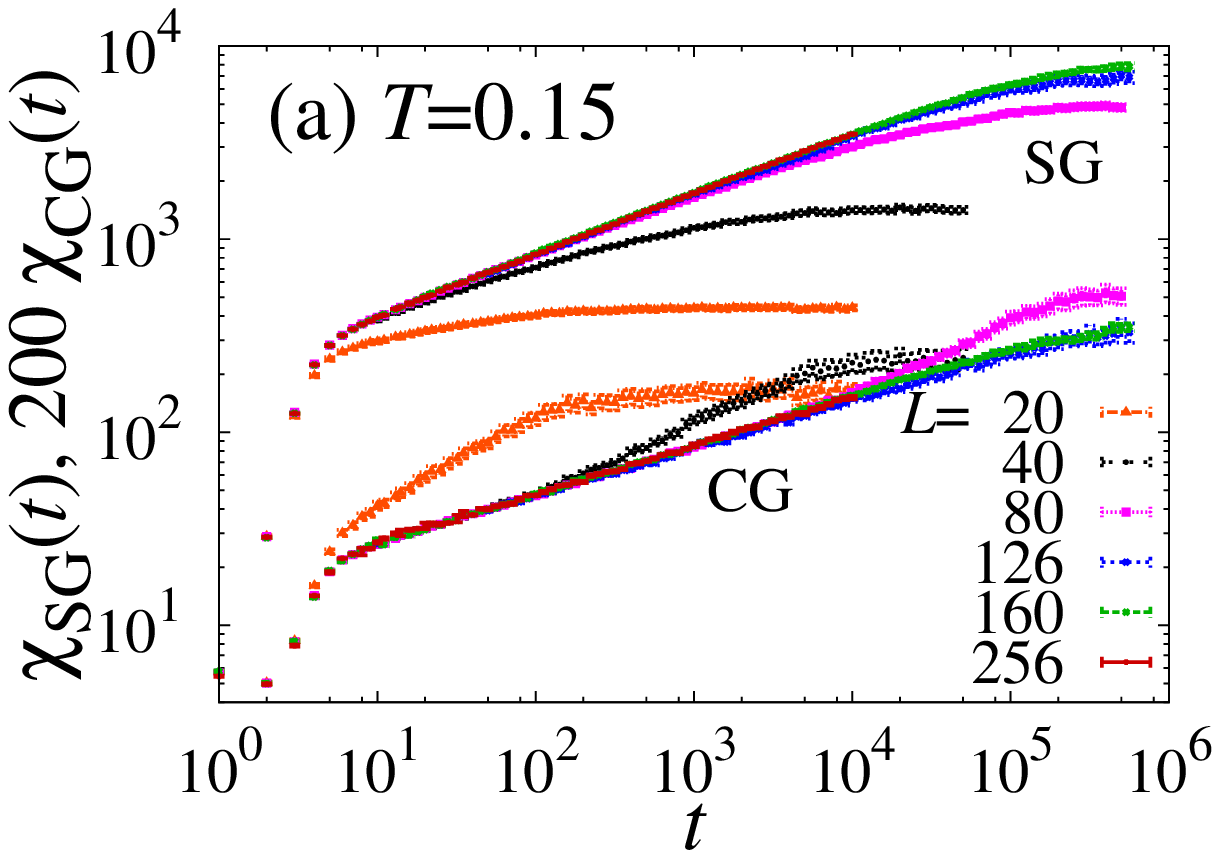}}
  \resizebox{0.40\textwidth}{!}{\includegraphics{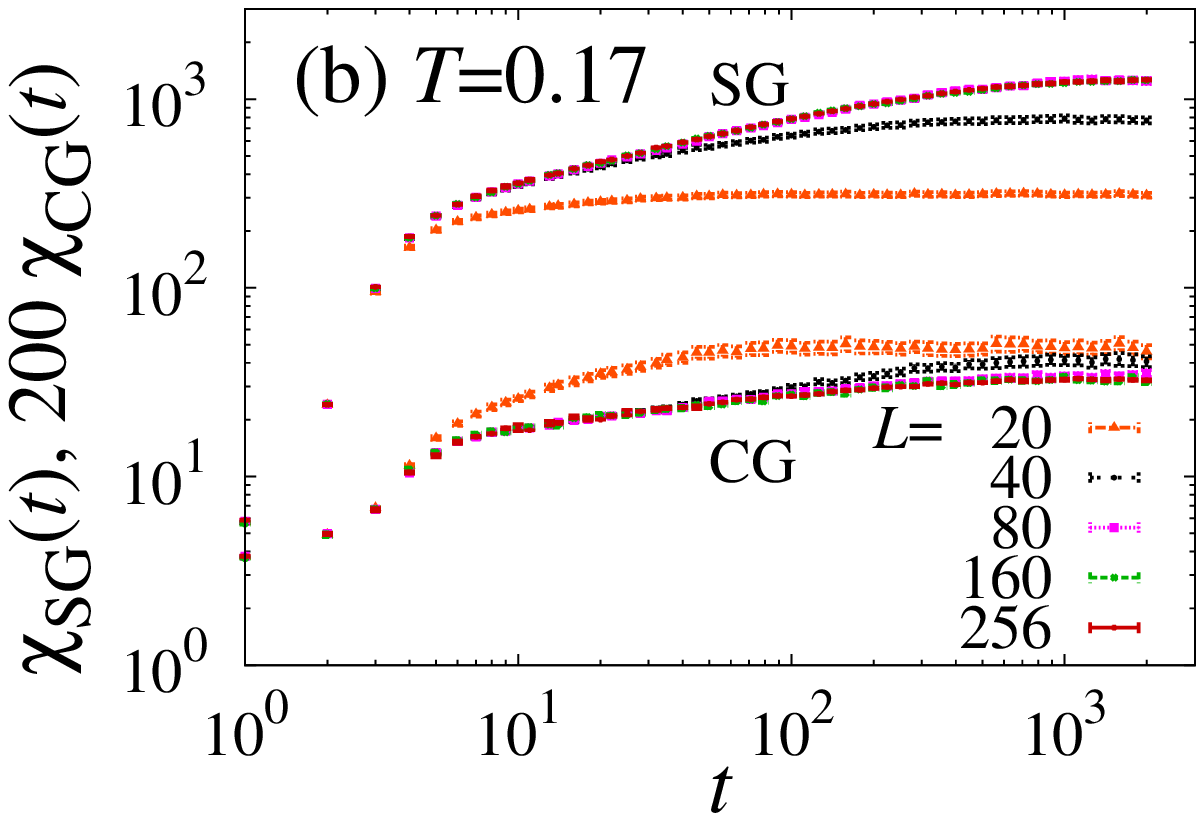}}
  \end{center}
  \caption{
(Color online)
Relaxation functions of $\chi_{\rm SG}$ and $\chi_{\rm CG}$ 
for various lattice sizes for (a) $T=0.15$ and (b) $T=0.17$.
Data of $\chi_{\rm CG}$ are multiplied by 200.
}
\label{fig:namatsize}
\end{figure}

We first check finite-size effects of $\chi_{\rm SG}$ and $\chi_{\rm CG}$.
Figure \ref{fig:namatsize} shows the relaxation functions for lattice sizes
from $L=20$ to 256 at $T=0.15$ and at $T=0.17$.
The temperatures are located in the paramagnetic phase.
We found a size-independent relaxation process and 
a size-dependent one in these figures.
The former one is regarded same as that of the infinite-size system. 
Relaxation functions of lattice sizes larger than 40 at $T=0.17$
show this behavior.
These simulations realize at the final step 
an equilibrium state in the thermodynamic limit.
The lattice sizes were large enough to contain an SG-ordered cluster.
On the other hand,
a relaxation function of $\chi_{\rm SG}$
deviated to a lower side, and that of $\chi_{\rm CG}$
deviated to an upper side,
when the lattice size is small.
For example, a relaxation function of 
$\chi_{\rm SG}$ for $L=40$ at $T=0.15$ started deviating when $t\simeq 10$.
This is a crossover time when the finite-size effects appeared.
The SG cluster is considered to reach a lattice boundary 
at this time step.
As the system size increases, this crossover occurred at later steps.
When the SG cluster size is smaller than the lattice size, 
relaxation functions do not exhibit size dependences.
A crossover of $\chi_{\rm CG}$ always occurred after that of $\chi_{\rm SG}$ 
occurred.
We consider that it is simply because the CG cluster is smaller
than the SG cluster.

Even though
the SG crossover of $L=40$ occurred at $t\simeq 10$,
it took $10^4$ steps to reach the equilibrium state.
Most of the time steps required for equilibration
were spent after this size effect appeared.
Why does it take such a long step?
We consider that the SG order is incompatible 
with the periodic boundary conditions.
The SG order tried to find another state that is
compatible with the boundary condition in this relaxation process.
A negotiation 
between the SG order and the boundary conditions took a very long time.
This is a slow dynamics observed in the equilibrium simulations.
We also found that the equilibration time steps of $\chi_{\rm SG}$ are
always equal to those of $\chi_{\rm CG}$ even though the finite-size
crossover times are different.
The SG order is waiting for the CG order to be equilibrated.

The finite-size effects of $\chi_{\rm SG}$ and $\chi_{\rm CG}$ are 
better understood by observing their profiles.
A profile of the susceptibility is a correlation function multiplied by 
$4\pi r^2$ plotted against $r$.
An integration of this value with respect to $r$ gives the susceptibility:
$\chi = \int_0^{L/2} 4\pi r^2 f(r) dr$, when $L$ is large enough.
We find by this plot 
how each correlation function contributes to the susceptibility, and
how the finite-size effect appears.
We can also estimate an effective size of the ordered cluster by a shape of 
this profile.

\begin{figure}
  \begin{center}
  \resizebox{0.35\textwidth}{!}{\includegraphics{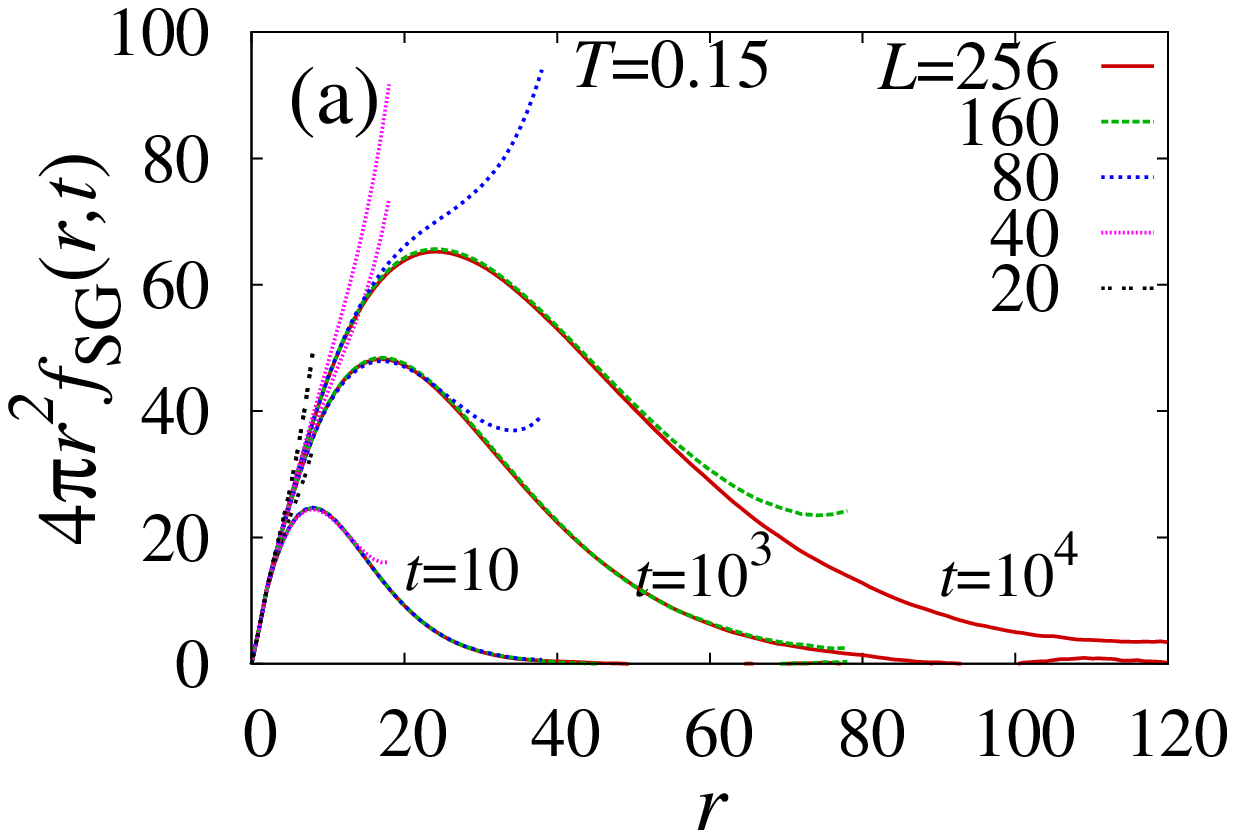}}
  \resizebox{0.35\textwidth}{!}{\includegraphics{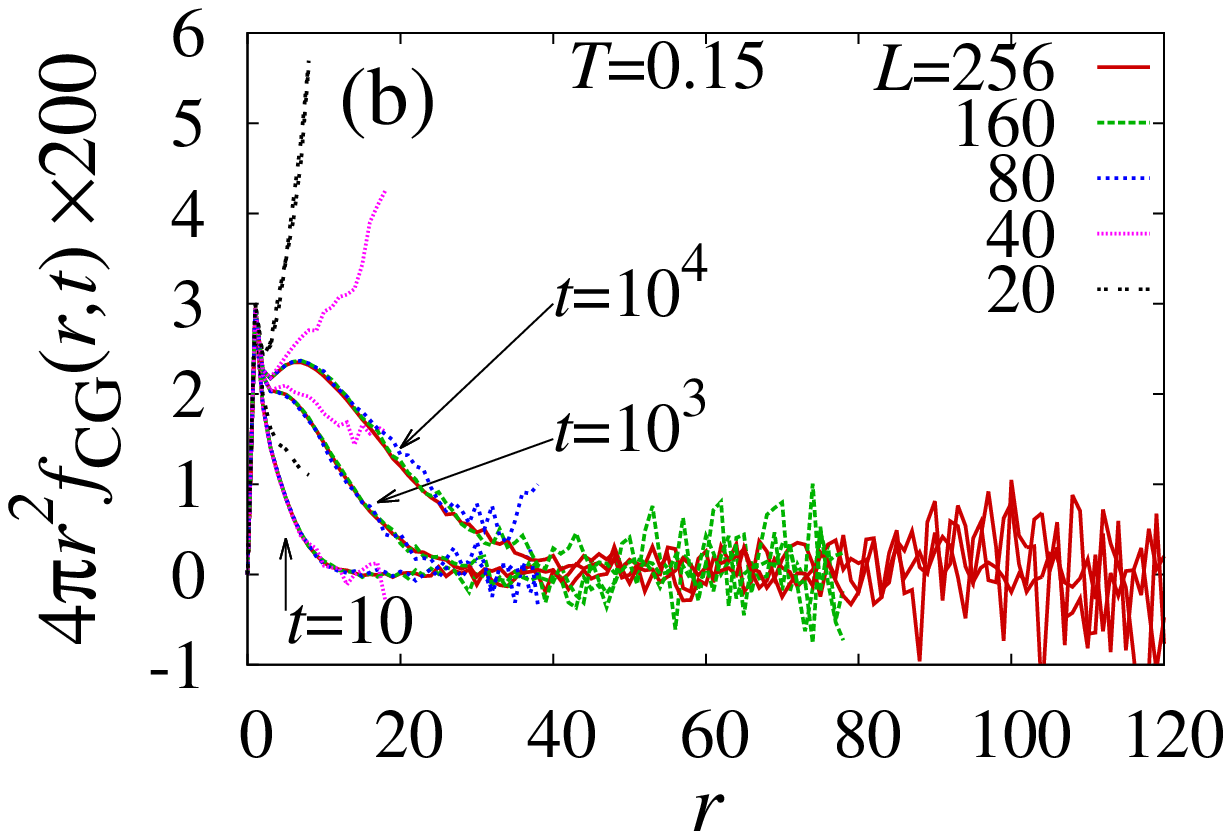}}
  \resizebox{0.35\textwidth}{!}{\includegraphics{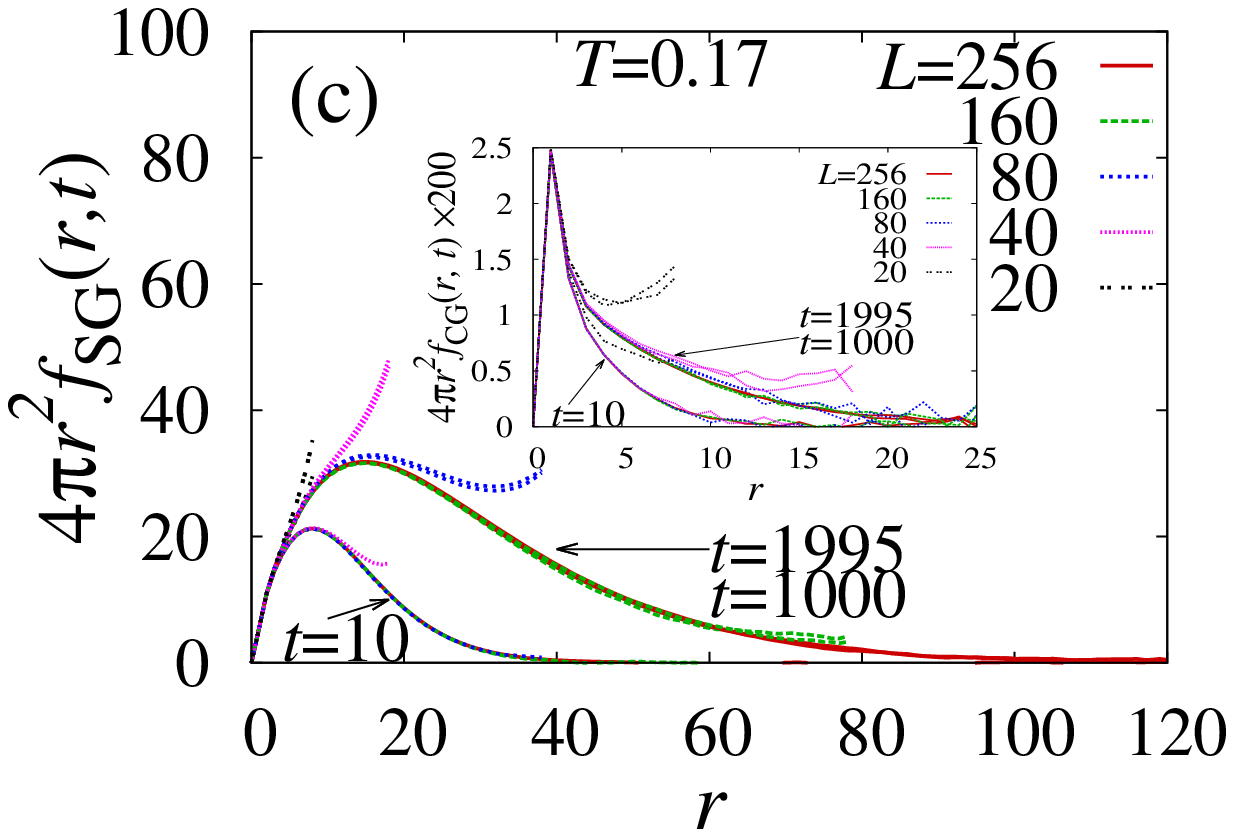}}
  \end{center}
  \caption{
(Color online)
 Profiles of $\chi_{\rm SG}$(a) and $\chi_{\rm CG}$(b) at $T=0.15$.
(c)
 Profiles of $\chi_{\rm SG}$ and $\chi_{\rm CG}$ (inset) at $T=0.17$.
}
\label{fig:profile}
\end{figure}

Figure \ref{fig:profile}(a) shows a profile of $\chi_{\rm SG}$ at 
$t=10$ , $10^3$, and $10^4$ for various lattice sizes when $T=0.15$.
They correspond to relaxation functions of $\chi_{\rm SG}$ 
in Fig.~\ref{fig:namatsize}(a).
The profiles exhibit a size-independent shape as long as a
cluster size did not exceed a lattice size.
A distance $r$ at which the profile line reaches zero is regarded as 
a radius of the ordered cluster.
Thus, we may regard its diameter, $2r$, as a size of the cluster.
The SG cluster size exceeded 60 even when $t=10$. 
The finite-size crossover of $\chi_{\rm SG}$ for $L=40$ beginning 
at $t\simeq 10$ is explained by this profile.
Data of lattice sizes larger than 40 traced on the same profile line,
while those of smaller sizes deviated upward.
After the SG cluster size reached the boundary, the SG correlation connects 
with each other beyond the periodic boundary.
The profile line is lifted due to this self correlation.
Finally in the equilibrium state of small lattices,
the profiles just exhibit monotonic increasing behaviors.
On the other hand, a profile of a larger lattice exhibits a long tail 
converging to zero.
Their contributions to the susceptibility are much larger than the 
ones from a monotonic-increasing profile of a smaller lattice.
Therefore, the SG susceptibility is always very much underestimated 
when a lattice size is small.

Figure \ref{fig:profile}(b) shows the profile of $\chi_{\rm CG}$ in the
same conditions of Fig.~\ref{fig:profile}(a).
A tail of profile drops rapidly even when a lattice size is large.
The CG cluster size is roughly three times smaller than that of SG.
There is an additional strong peak at $r=1$.
It is explained by a definition of a chirality,
which is a product of three neighboring spins.
The peak at $r=1$ is an outcome of a self-correlation of chirality.
It causes a strong finite-size enhancement when a lattice size is small.
Therefore, a finite-size effect of $\chi_{\rm CG}$ always appears
as overestimating.

Figure \ref{fig:profile}(c) shows profiles of $\chi_{\rm SG}$ and 
$\chi_{\rm CG}$ when $T=0.17$.
They correspond to a relaxation functions in Fig.~\ref{fig:namatsize}(b).
Profiles when $t=10$ are same as those at $T=0.15$.
This short-time behavior is almost independent of the temperature.
On the other hand,
profiles of $L \ge 160$ when $t=1000$ and $t=1995$
show no size dependence. 
The system reached the equilibrium state at these time steps.
The profiles are considered as those in the thermodynamic limit at this
temperature.
A shape of the equilibrium profile is qualitatively same as those in the 
nonequilibrium process before the finite-size effects appeared.

 \begin{figure}
   \begin{center}
   \resizebox{0.40\textwidth}{!}{\includegraphics{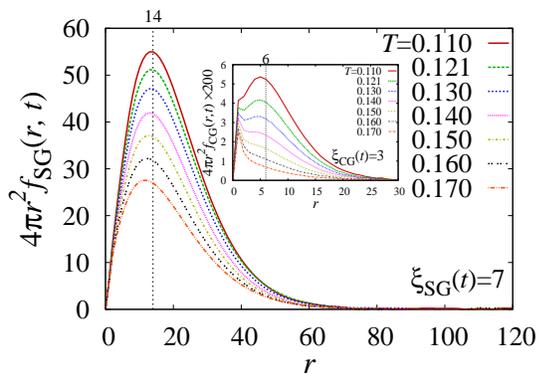}}
   \end{center}
   \caption{
 (Color online)
 Profiles of $\chi_{\rm SG}$ at a step when $\xi_{\rm SG}(t)=7$ 
 for various temperatures.
 Inset shows profiles of $\chi_{\rm CG}$ multiplied by 200 
 at a step when $\xi_{\rm CG}(t)=3$.
 The linear lattice size is 256.
 }
 \label{fig:nbehavior}
 \end{figure}

We also studied profiles before the finite-size effects appear.
Figures \ref{fig:nbehavior} shows profiles of $\chi_{\rm SG}$ and 
$\chi_{\rm CG}$ at a time step when $\xi_{\rm SG}(t)=7$ and 
$\xi_{\rm CG}(t)=3$ for various temperatures.
As the temperature decreases, an amplitude of profile grows 
and the peak position approaches $2\xi$, while keeping the shape.
Profiles of $\chi_{\rm CG}$ show similar behaviors, but
it has an additional sharp peak at $r=1$.

\begin{figure}
  \begin{center}
  \resizebox{0.40\textwidth}{!}{\includegraphics{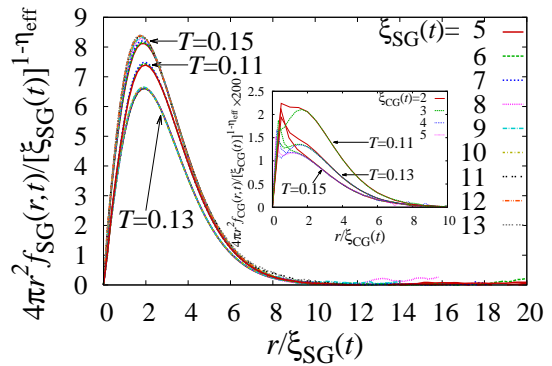}}
  \end{center}
  \caption{
(Color online)
Profiles of scaled $\chi_{\rm SG}$ 
plotted against $r/\xi_{\rm SG}$
when $\xi_{\rm SG}$ reached the denoted value 
ranging from 5 to 13.
The effective exponents are 
$\eta_{\rm eff}=-0.030$ for $T=0.11$,
$\eta_{\rm eff}=-0.0087$ for $T=0.13$,
and
$\eta_{\rm eff}=0.22$ for $T=0.15$.
Inset is a same plot for $\chi_{\rm CG}$ multiplied by 200.
The effective exponents are 
$\eta_{\rm eff}=0.15$ for $T=0.11$,
$\eta_{\rm eff}=0.19$ for $T=0.13$,
and
$\eta_{\rm eff}=0.59$ for $T=0.15$.
The linear lattice size is 256.
}
\label{fig:nbehavior2}
\end{figure}

Figure \ref{fig:nbehavior2} shows the scaled profiles
at a time step when the correlation length
reached each value ranging from 5 to 13 for SG and 
that ranging from 2 to 5 for CG.
Since 
$\chi_{\rm SG}\sim \xi_{\rm SG}^{2-\eta}$,
a profile of $\chi_{\rm SG}$ is scaled by $\xi_{\rm SG}^{1-\eta}$ if
plotted against $r/\xi_{\rm SG}$.
Here, $\eta$ is an effective exponent obtained by the correlation-function
scaling when we estimated the correlation length.
A shape of the scaled profile remains the same at each temperature.
It has a peak at $r/\xi\simeq 2$.
This is because
the correlation functions exhibit a single-exponential decay as 
$f(r)\sim \exp[-r/\xi]$.
A collapse of the scaling became poor at $T=0.15$.
This temperature is higher than the critical temperature 
and the scaling hypothesis may not be well satisfied.

We found in these figures
that the SG and CG profiles always reach zero when $r/\xi > 10$.
We can guarantee that the finite-size effects do not appear
if we set $L>2r=20~\xi_{\rm SG}$.
This is a criterion of choosing lattice size and the
simulation time range in this paper.

\subsection{Sample dependences}

We must take averages of physical quantities over different 
random samples in SG simulations.
Collected data are considered to depend on each sample.
Before taking this sample average, we must take the thermal average.
In an equilibrium SG simulation scheme, the thermal average has been performed 
by the MC time average using two real replicas.
In this paper, we study the SG phase transition by the relaxation 
functions of physical quantities.
We need at each step a value after taking the thermal average.
Therefore, we introduced an average over real replicas 
as the thermal average.\cite{nakamura}
We must choose a large replica number for a better accuracy.
Then, a sample number, $n_{\rm s}$, is restricted,
because
a total computational time is roughly proportional to $L^3 mn_{\rm s}$.
So, there arises a question.
Which number should be set large first, a replica number $m$ or
a sample number $n_\mathrm{s}$?

\begin{figure}
  \begin{center}
  \resizebox{0.35\textwidth}{!}{\includegraphics{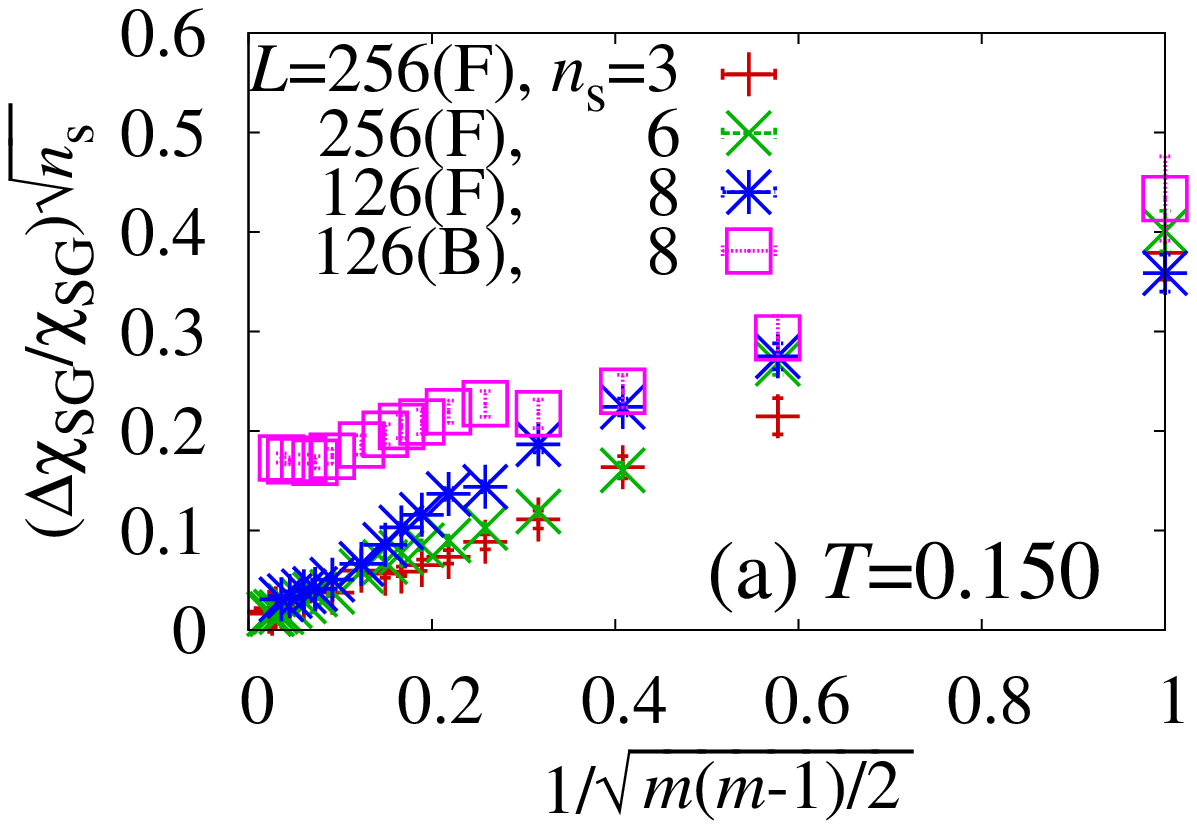}}
  \resizebox{0.35\textwidth}{!}{\includegraphics{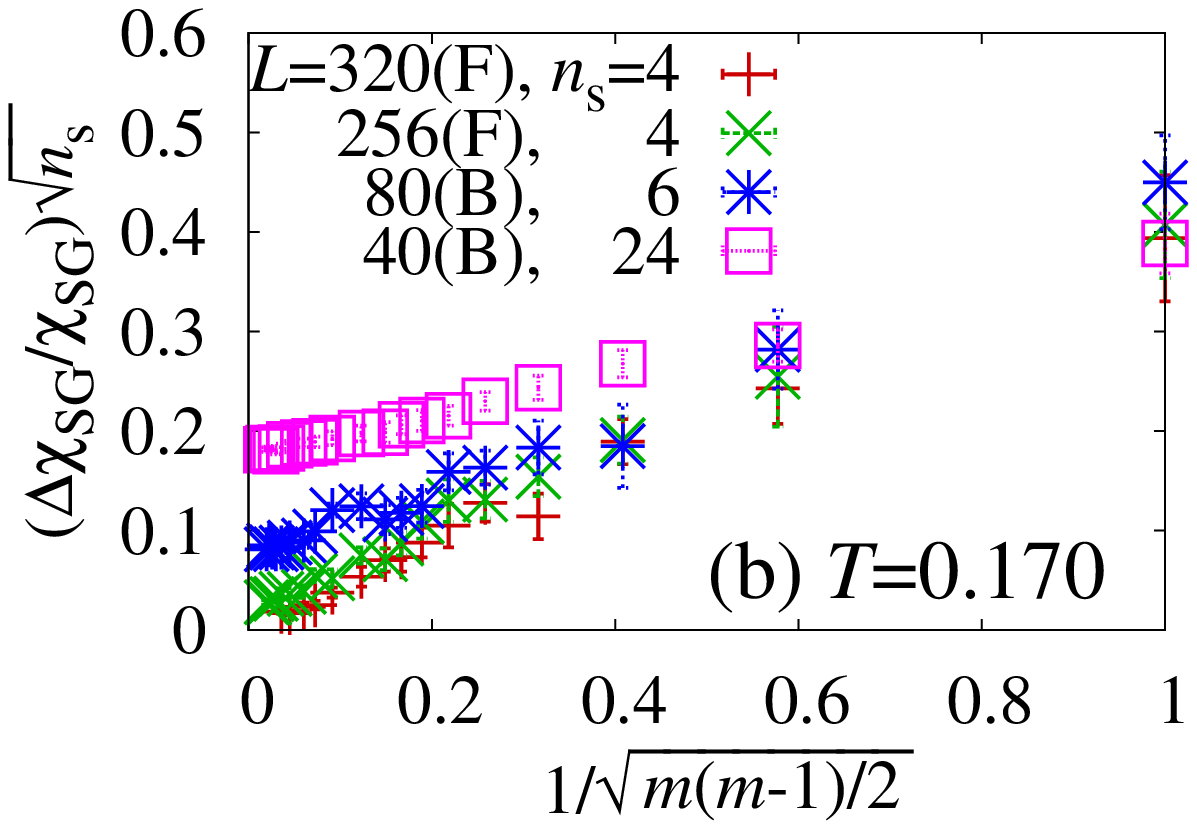}}
  \resizebox{0.35\textwidth}{!}{\includegraphics{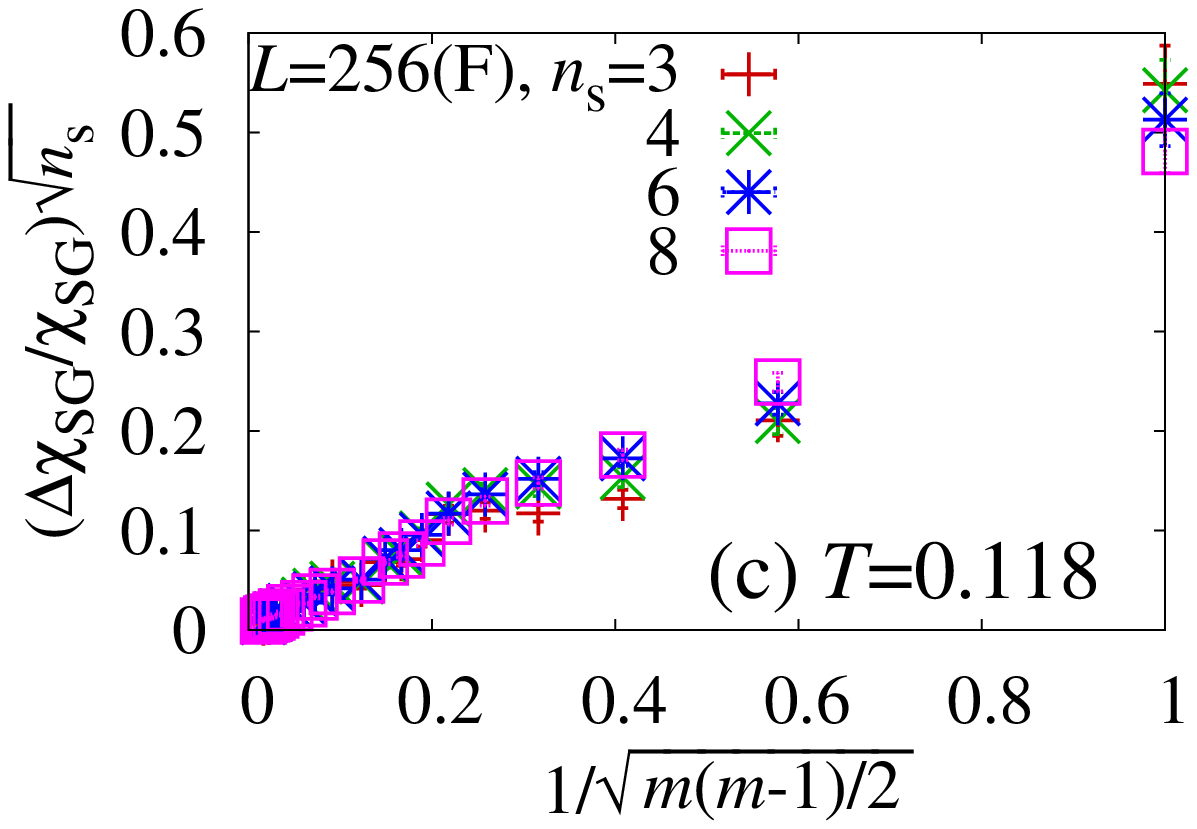}}
  \end{center}
  \caption{
(Color online)
Relative errors of $\chi_{\rm SG}$ multiplied by $\sqrt{n_{\rm s}}$ 
are plotted against $1/\sqrt{m(m-1)/2}$ for various
lattice sizes and sample numbers for (a)$T=0.15$, (b) $T=0.17$, and
(c) $T=0.118$.
(F) stands for the data taken when the $\chi_{\rm SG}$ profile 
was (F)ree from the the boundary.
(B) stands for the data taken after the $\chi_{\rm SG}$ profile
reached the (B)oundary.
}
\label{fig:selfave}
\end{figure}

Figure~\ref{fig:selfave}(a) shows the answer.
We estimated a relative error of $\chi_{\rm SG}$ 
and multiplied it by $\sqrt{n_{\rm s}}$.
It is a standard deviation.
They are plotted against $1/\sqrt{m(m-1)/2}$. 
We changed a replica number from 2 to 72, and a sample number from 3 to 8.
We also compared data free from the boundary effects 
and those affected by them.
Data free from the boundary effects
were taken in the nonequilibrium process before the SG cluster
size reached the lattice boundary.
They rode on a straight line as a replica number increases, and
converged to zero in a limit of $m\to\infty$.
We also found that the relative errors are proportional to 
$1/\sqrt{m(m-1)/2}$ not to $1/\sqrt{m}$.
This suggests that each replica {\it overlap} is independent, not that
each replica is independent.
Data affected by the boundary effects were taken in the nonequilibrium
process after the SG cluster size reached the lattice boundary.
They converged to a finite value.

Figure~\ref{fig:selfave}(b)
shows data that were estimated after the equilibrium states were realized 
at $T=0.17$. 
An equilibrium cluster size is roughly 200 as shown in
Fig.~\ref{fig:profile}(c).
Data of $L>200$ are free from the boundary effect, and converged linearly
to zero.
Those of $L<200$ converged to finite values, which decrease as $L$ increases.
Figure~\ref{fig:selfave}(c) shows data below the critical temperature.
The SG cluster size did not exceed the lattice size within the 
simulated time steps.
They also converged linearly to zero.
These data exhibit the same tendency as those of nonequilibrium process 
at $T=0.15$.
Therefore,
data of each random sample are considered to be
independent and {\it equivalent} in a limit of $m\to\infty$, 
if the profiles are free from the 
boundary effect no matter whether it is in the equilibrium state
or in the nonequilibrium state, and also no matter whether the
temperature is above or below the critical temperature.

Since the computational cost is proportional to $L^3mn_{\rm s}$, 
it is better to increase $m$ first 
in order to reduce the numerical errors within a restricted computational time.
In this paper, we set $m$ to 64 or 72, and set 
$n_{\rm s}$  mostly to 4-8 when $L=256$.
We increased $n_{\rm s}$ up to 10 according to the sample fluctuations 
particularly near the critical temperature.

\section{Our strategy}
\label{sec:method}

Difficulties in SG simulations are strong finite-size effects,
strong sample dependences, and the slow dynamics.
In the previous section, we found that a competition between the SG
order and the boundary condition is the 
main origin of these difficulties.

The first step of our strategy is to remove the size effects by using a
large lattice size, $L > 20~\xi_{\rm SG}$.
The second one is to solve the sample dependence 
by increasing a replica number.
The final difficulty is the slow dynamics.
We solve it by giving up the equilibrium simulation, and
study the relaxation functions of physical quantities.
The nonequilibrium relaxation method\cite{Stauffer,Ito,nerreview} 
realizes this strategy.
Together with the dynamic correlation-length scaling method,\cite{totasca} 
we clarify in this paper
the unsettled issue of the SG and the CG phase transition
in the Heisenberg SG model in three dimensions.
We consider that this strategy is justified because the SG 
cluster size is so large even within the short MC time steps.
Let us briefly explain our methods in the followings.

The nonequilibrium relaxation method\cite{Stauffer,Ito,nerreview}
studies a phase transition through 
the relaxation functions of physical quantities.
We run a simulation on a very large system and stop the simulation before
the finite-size effect appears.
Thus, the obtained relaxation functions are regarded 
as those of the infinite-size system.
We can determine the critical temperature and critical exponents 
by the dynamic (finite-time) scaling analysis.
Since the system size is regarded as infinite, this method is successfully 
applied\cite{OzIto,OzIto2,shirahata1,nakamura2,shirahata2,yamamoto1,%
nakamura4l,nakamura4}
to systems with frustration and randomness, which causes strong size effects.

In the SG simulations,
we simulate $m$ independent real replica systems for one bond sample
starting with independent initial spin states.
The thermal averages are taken over real replicas at each observation time.
Then, we obtain relaxation functions of 
physical quantities for one bond sample.
Changing the initial spin state, the random bond sample, 
and the random number sequence, we start
another set of simulations to obtain another set of the relaxation functions.
We calculate the average of the relaxation functions
over the random bond samples.
Here, we note that every average procedure is taken over independent data.

A scaling analysis is based on the scaling hypothesis,
\begin{equation}
 \chi \sim \xi^{2-\eta}, ~~~ \xi \sim |T-T_{\rm c}|^{-\nu}.
\label{eq:scahyp}
\end  {equation}
The critical temperature is denoted by $T_{\rm c}$ in this expression.
In the finite-size-scaling analysis, 
we replace $\xi$ by $L$ in Eq.~(\ref{eq:scahyp}) supposing
$L$-$\xi$ equivalence in the scaling region.
By using {\it equilibrium} data of the susceptibility 
$\chi(L, T)$ for each $L$ and $T$,
$\chi(L, T)/L^{2-\eta}$ data are plotted against $L/|T-T_{\rm c}|^{-\nu}$.
We determine $T_{\rm c}$, $\nu$, and $\eta$ 
so that the scaled data ride on a single curve.
In the finite-time-scaling analysis of the nonequilibrium relaxation 
method,\cite{nerreview}
we replace $\xi$ by $t^{1/z}$ in Eq.~(\ref{eq:scahyp}), where
$z$ is a dynamic exponent.
This replacement is guaranteed by the dynamic scaling hypothesis, 
$t \sim  \xi^z$.
Using a {\it nonequilibrium} 
relaxation function of $\chi$ for various temperatures, we plot
$\chi(t, T)/t^{(2-\eta)/z}$ against $t/|T-T_{\rm c}|^{-z\nu}$
so that the scaled data ride on a single curve.
We can obtain $T_{\rm c}$, $z\nu$, and $\gamma (=\nu\times (2-\eta))$
by this scaling plot.

In this paper,
we investigate the critical phenomena using the
dynamic correlation-length scaling analysis.\cite{totasca}
This is a direct application of the scaling hypothesis
to the nonequilibrium relaxation data.
In this analysis, we replace $\xi$ by its relaxation function $\xi(t, T)$, 
and replace $\chi$ by its relaxation function $\chi(t, T)$ 
in Eq.~(\ref{eq:scahyp}).
We plot
$\chi(t, T)/\xi^{2-\eta}(t, T)$ against $\xi(t, T)/|T-T_{\rm c}|^{-\nu}$ 
and estimate $T_{\rm c}$, $\nu$, and $2-\eta$ so that 
all the data fall on a single curve.
This estimation is performed using the Bayesian inference proposed by
Harada.\cite{harada}
It realizes unbiased and precise estimations of critical parameters.

One advantage of the dynamic correlation-length scaling analysis 
is that both finite time, $t$, 
and  finite size, $L$, do not appear explicitly in the scaling expression.
We only deal with the physical quantities, $\chi$ and $\xi$. 
Usually, a finite size and a finite time produce nontrivial effects
in the SG system, and probably in general complex systems.
Scaling analyses replacing $\xi$ by size or time may need 
special attentions to the scaling form we treat.
Additional correction-to-scaling terms are sometimes necessary.
Such nontrivial effects become hidden in the present 
correlation-length scaling analysis.
Nontrivial time dependences of $\xi(t)$ and $\chi(t)$ can be cancelled
if we plot $\chi(t)$ against $\xi(t)$.

Let us summarize our simulation conditions here.
MC simulations are performed by the single-spin-flip algorithm.
One MC step consists of 
one heat-bath update, 124 over-relaxation updates, and
1/20 Metropolis update(once every 20 steps).
We start simulations with random spin configurations.
The temperature is quenched to a finite value at the first Monte Carlo step.
The linear lattice size was fixed to 256.
The temperature ranges from $T=0.02$ to $T=0.18$ at 73 different 
temperature points.
Random bond configurations are generated independently at each temperature.
The sample numbers are mostly 6, but we increased it up to 10 when the
data fluctuations were large.
Total sample number for all the temperatures is 432.
A replica number is mostly 72.
We increased it to 88 at some temperatures in order to check 
if there are systematic dependences on a replica number.
In the scaling analysis, we discarded data at very low temperatures,
$T<0.10$, because the scaled data separate from the data of $T\ge 0.10$.
A typical initial step is 50, and a typical final step is 10000.
We increased it at most up to 31623 at low temperatures.
Only data with $\xi_{\rm SG}(t)< L/20=12.8$ are used in the scaling analysis.

\section{Results}
\label{sec:results}

\begin{figure}
   \resizebox{0.35\textwidth}{!}{\includegraphics{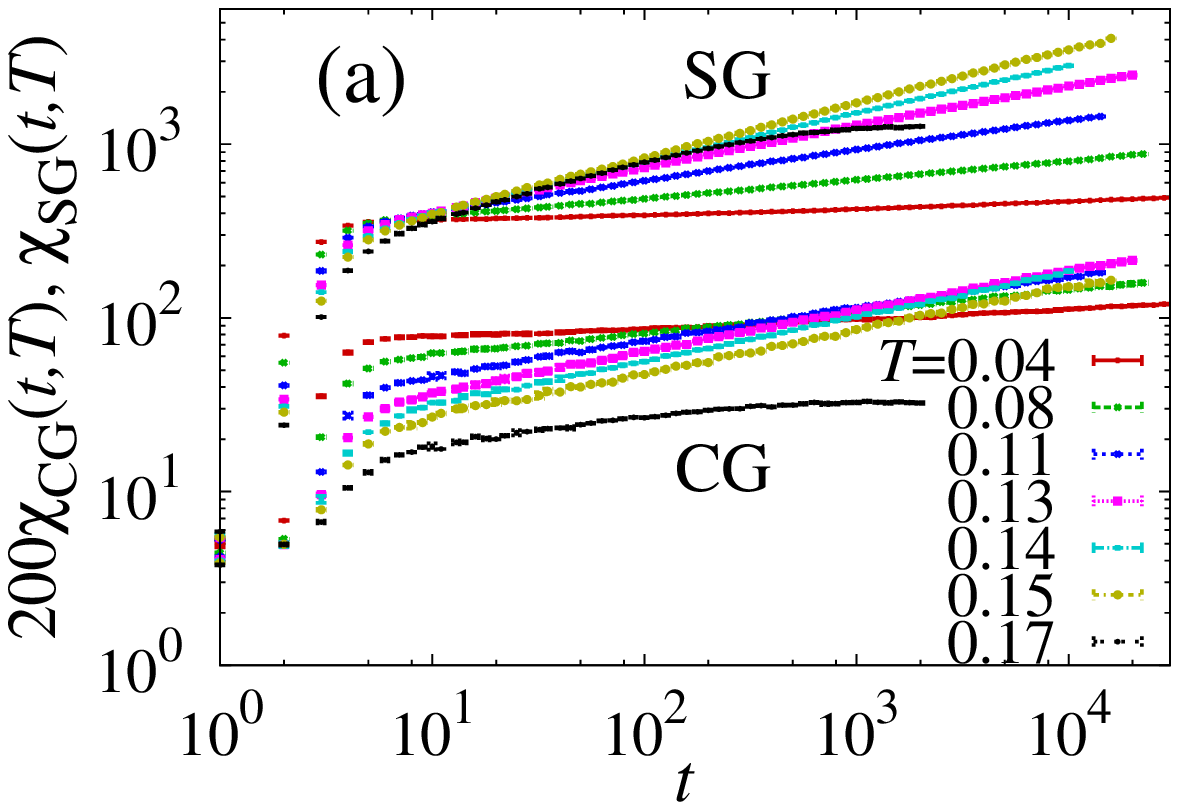}}
   \resizebox{0.35\textwidth}{!}{\includegraphics{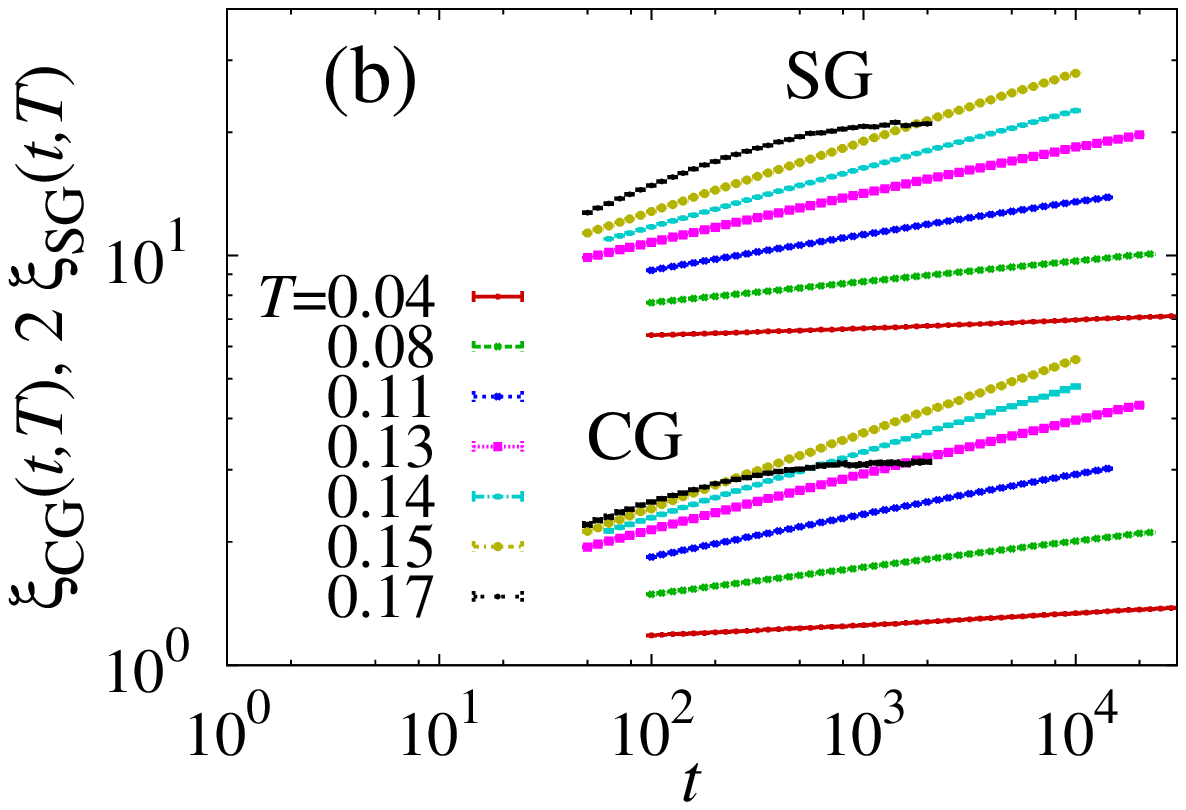}}
   \resizebox{0.35\textwidth}{!}{\includegraphics{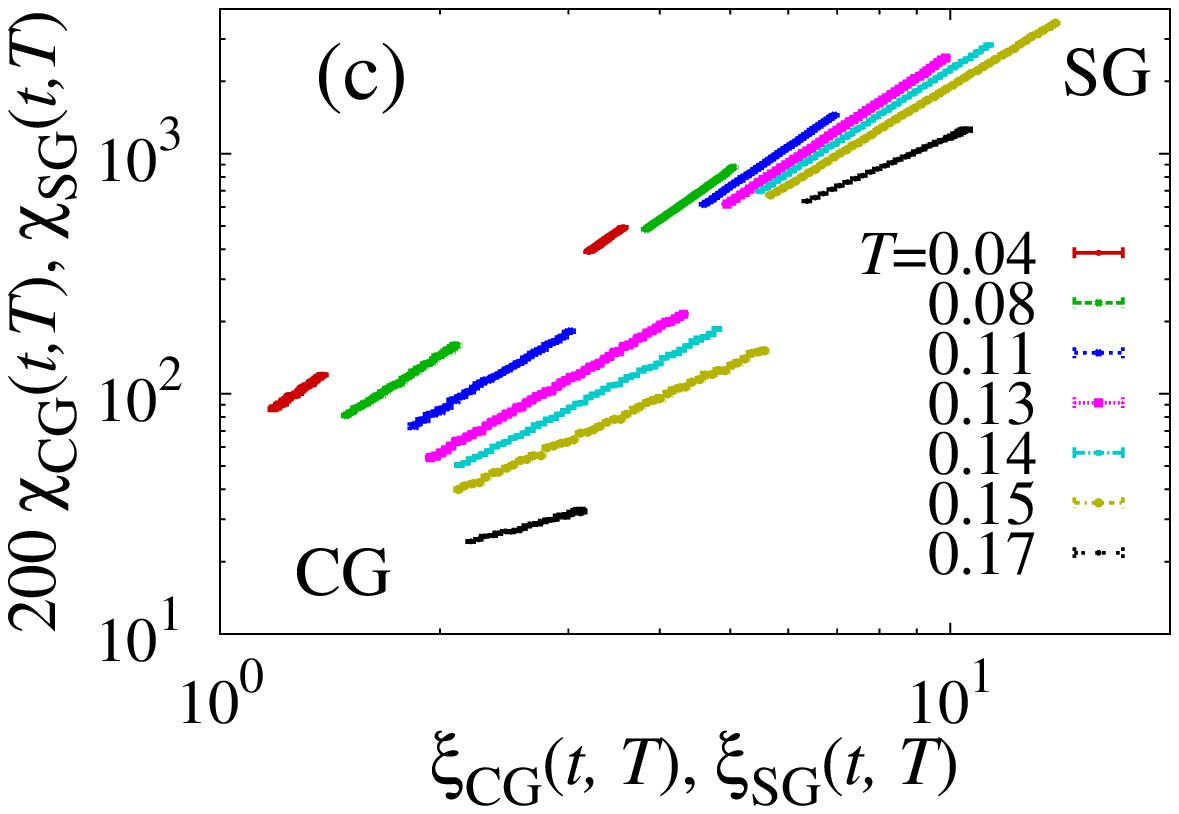}}
   \caption{
 (Color online)
(a)
 Relaxation functions of $\chi_{\rm SG}$ and $\chi_{\rm CG}$.
 Data of $\chi_{\rm CG}$ are multiplied by 200 in order to fit in 
 a same window.
(b)
 Relaxation functions of $\xi_{\rm SG}$ and $\xi_{\rm CG}$.
 Data of $\xi_{\rm SG}$ are multiplied by 2 in order to separate them 
 from the $\xi_{\rm CG}$ data.
 (c) 
  A cross plot of the susceptibility against the correlation length.
 }
 \label{fig:namat}
 \end{figure}

Figure \ref{fig:namat}(a) shows relaxation functions of 
$\chi_{\rm SG}$ and $\chi_{\rm CG}$ at typical temperatures.
We found a change of relaxation behavior at $t\sim 10$.
Data before $t\sim 10$ are considered as in an initial relaxation process.
Both $\chi_{\rm SG}$ and $\chi_{\rm CG}$ rapidly increase
at lower temperatures.
A size of the SG cluster reached 80 lattice spacings as was
shown in Fig.~\ref{fig:profile}(a).
Data after $t\sim 10$ are considered as in the critical relaxation process.
They are relevant to the phase transition.
A slope of this figure corresponds to a ratio of critical exponents,
$(2-\eta)/z=\gamma / z\nu$. 
It decreases with the temperature decreasing reflecting an increase of
the dynamic exponent in the low-temperature phase.
Figure \ref{fig:namat}(b) shows the corresponding relaxation functions of
correlation lengths.
A slope of this figure is an inverse of the dynamic exponent: $1/z$.
We plotted $\chi(t)$ against $\xi(t)$ in Fig.~\ref{fig:namat}(c).
We found that there is no bending anomaly from  the nonequilibrium relaxation 
process to the equilibrium relaxation process.
This plot tells us that both processes smoothly connect with each other
if we plot $\chi(t)$ against $\xi(t)$.

Using $\chi(t, T)$ plotted against $\xi(t, T)$, we performed the 
dynamic correlation-length scaling analysis.
Then, we obtained the critical temperature and the critical exponents.
There were 2816 data points of $(\xi, \chi)$ 
for different time steps and temperatures.
We randomly selected 1400 data points out of them and applied the kernel
method to obtain the critical temperature and critical exponents
such that the selected data ride on the scaling function.
We checked the obtained results by a cross validation method. 
Namely, we randomly selected 1400 data points again and tested the 
obtained parameters by estimating a likelihood function, $\Lambda$.
We tried this check for ten times by changing the selected data and
took an average of $-\ln (\Lambda)$ over them.
Then, one estimated set of $(T_{\rm c}, \nu, 2-\eta)$ and $-\ln (\Lambda)$
are obtained.
We repeated this trial for 100 times and took averages over
results whose $-\ln(\Lambda)$ values only differ within the standard
deviation from the best value.
We put error bars by this standard deviation among these results.

\begin{figure}
  \resizebox{0.23\textwidth}{!}{\includegraphics{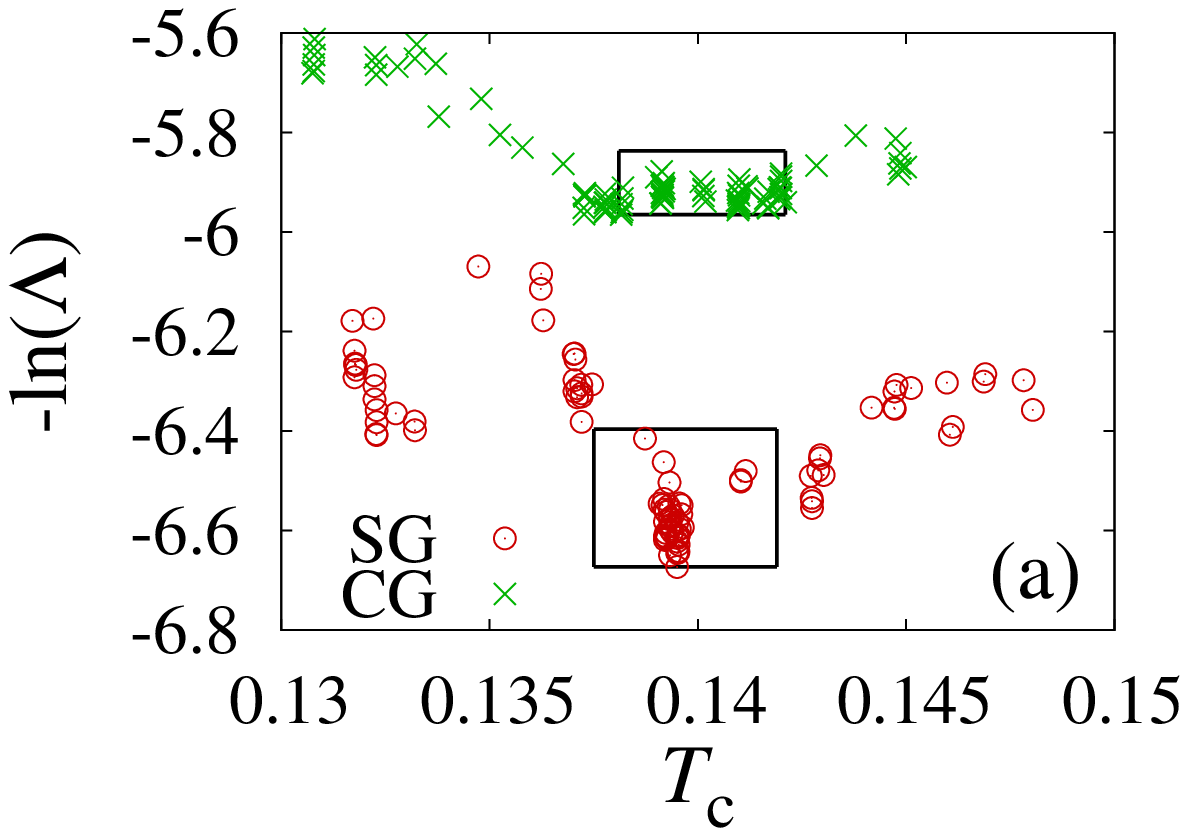}}
  \resizebox{0.23\textwidth}{!}{\includegraphics{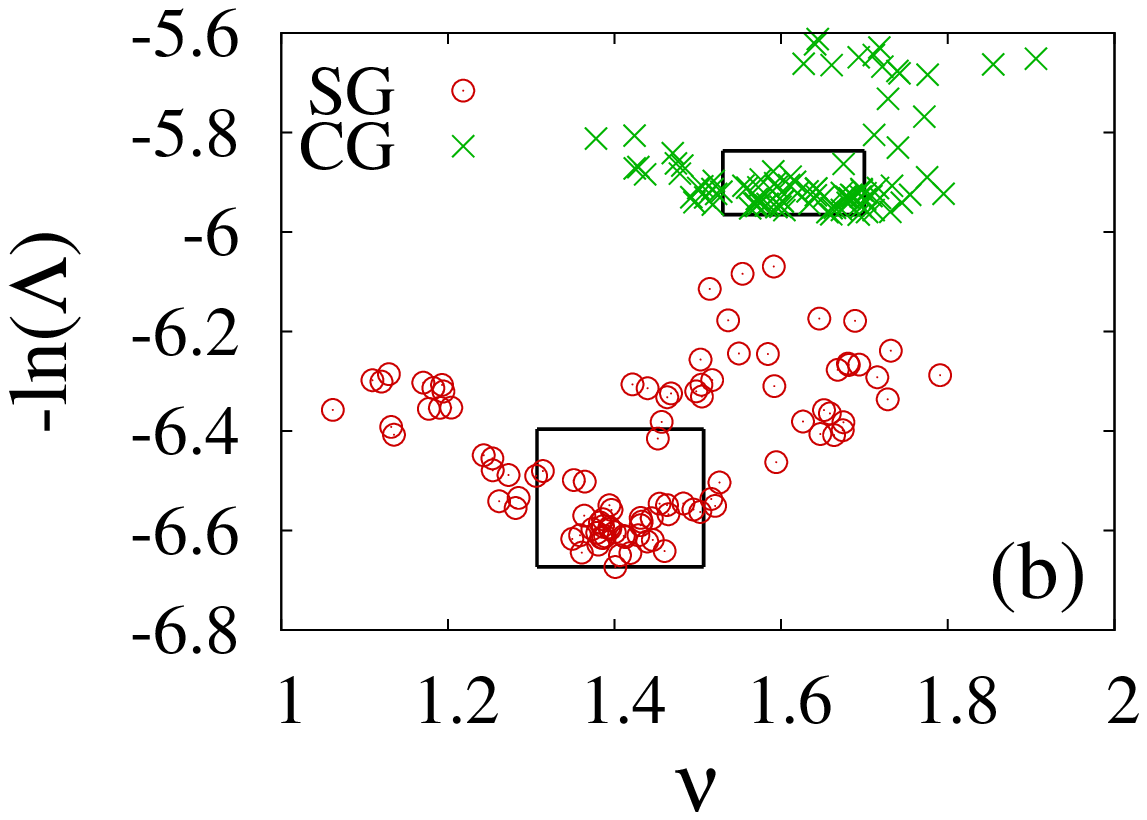}}
  \resizebox{0.23\textwidth}{!}{\includegraphics{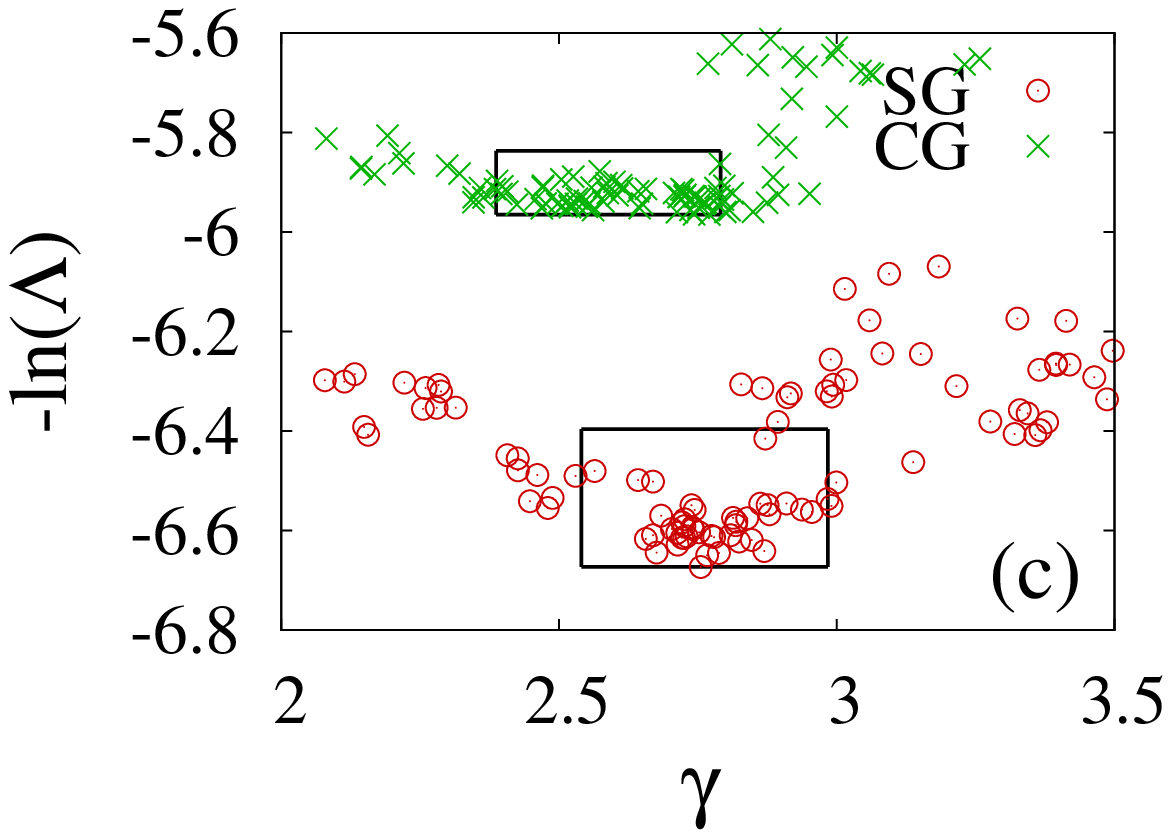}}
  \resizebox{0.23\textwidth}{!}{\includegraphics{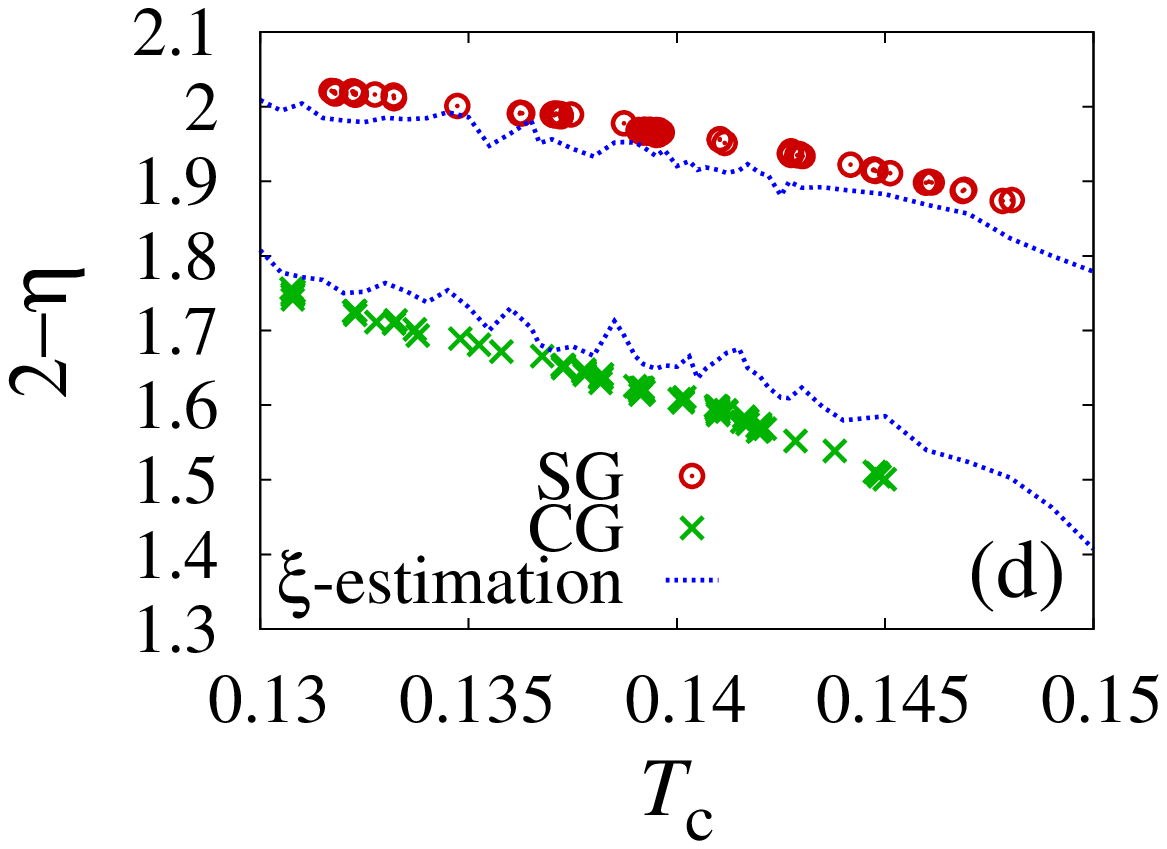}}
  \caption{
(Color online)
(a), (b), (c)
Likelihood functions plotted against the estimated
parameters for each scaling trial.
(d) Relation between the estimated $(2-\eta)$ and the critical temperature.
Effective $2-\eta$ obtained in the $\xi$ estimation is
also plotted with line.
}
\label{fig:sca-tc}
\end{figure}

Results of the trial are shown in Fig.~\ref{fig:sca-tc}.
Figures \ref{fig:sca-tc}(a)-(c) show the $-\ln(\Lambda)$ 
plotted against the estimated critical temperature,
the estimated $\nu$, and the estimated $\gamma(=\nu\times (2-\eta))$, 
respectively.
An estimate is better if $-\ln(\Lambda)$ is lower.
A rectangle shows the estimated error bar.
Figure \ref{fig:sca-tc}(d) shows relations between the estimated
$2-\eta$ and the estimated critical temperature.
We also plotted with lines the effective $2-\eta_{\rm eff}$ 
obtained in the $\xi$ estimation.
It is expected to coincide with $2-\eta$ at $T=T_{\rm c}$.
However, there are small differences between them.

\begin{figure}
  \resizebox{0.40\textwidth}{!}{\includegraphics{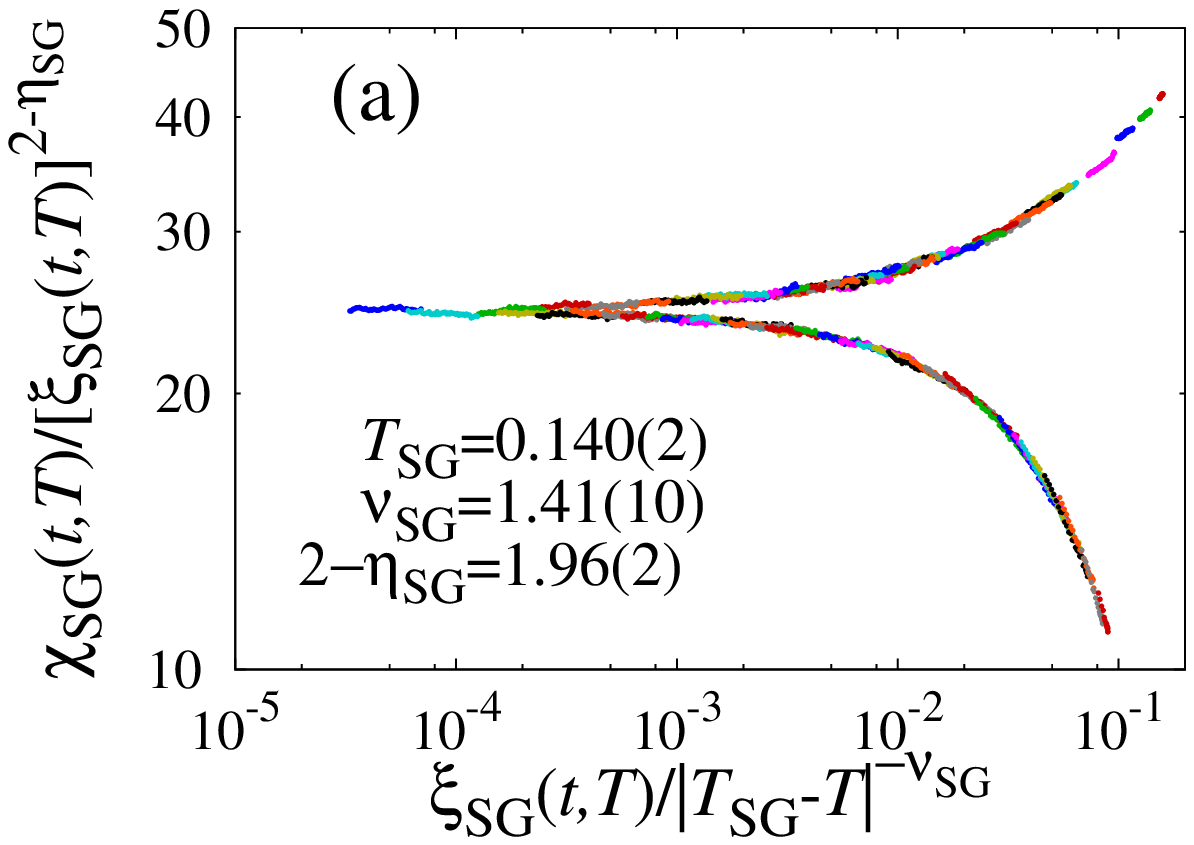}}
  \resizebox{0.40\textwidth}{!}{\includegraphics{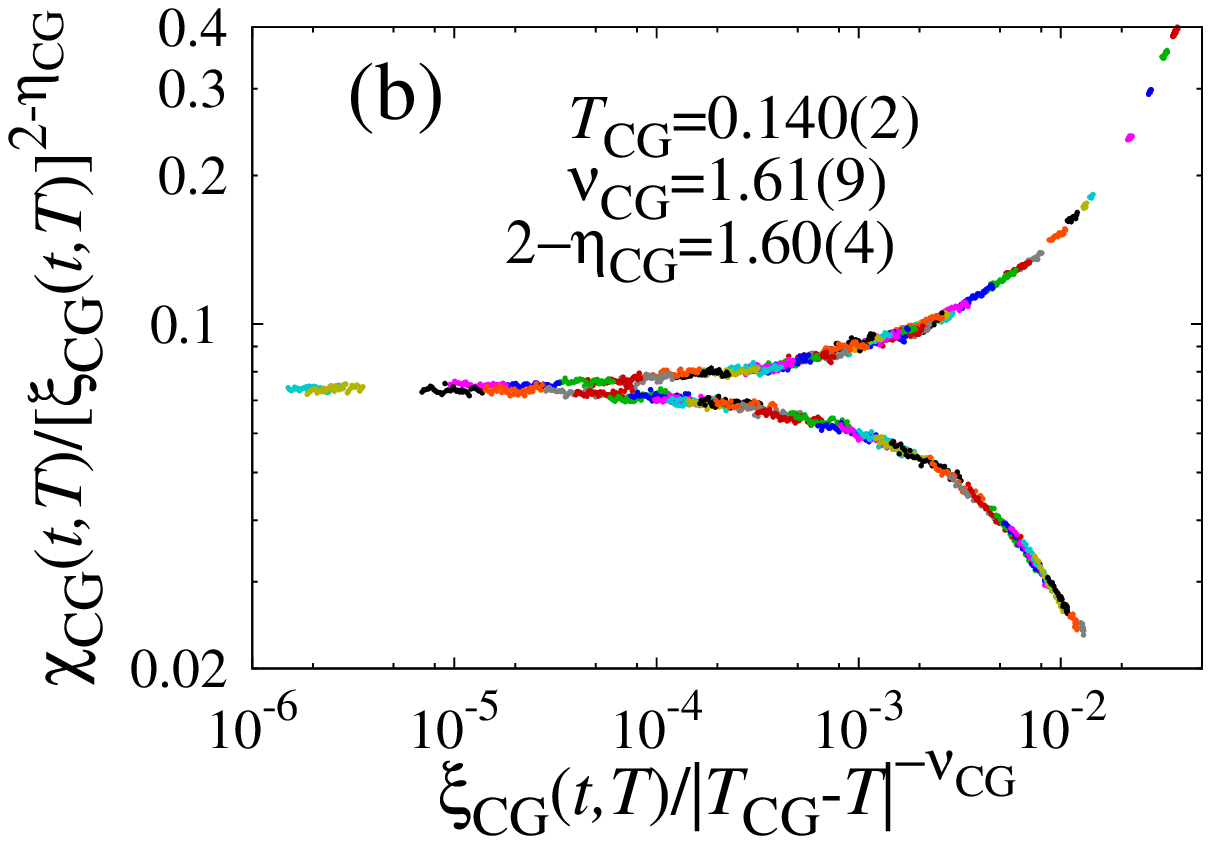}}
  \caption{
(Color online)
Dynamic correlation-length scaling plots of $\chi_{\rm SG}$-$\xi_{\rm SG}$(a)
and
$\chi_{\rm CG}$-$\xi_{\rm CG}$(b).
Data of $0.02\le T \le 0.18$ are plotted.
}
\label{fig:scalingplot}
\end{figure}

Figure \ref{fig:scalingplot} shows the scaling plot using the 
estimated critical parameters:
\begin{eqnarray}
T_\mathrm{SG} &=& 0.140 \pm 0.002~(0.1395),\\
\nu_\mathrm{SG} &=& 1.41 \pm 0.10~(1.401),\\
2-\eta_\mathrm{SG}&=&1.96\pm 0.02~(1.967),\\
\gamma_\mathrm{SG}&=&2.76\pm 0.22~(2.755),
\end  {eqnarray}
and 
\begin{eqnarray}
T_\mathrm{CG} &=& 0.140 \pm 0.002~(0.1382),\\
\nu_\mathrm{CG} &=& 1.61 \pm 0.09~(1.693),\\
2-\eta_\mathrm{CG}&=&1.60\pm 0.04~(1.637),\\
\gamma_\mathrm{CG}&=&2.59\pm 0.20~(2.771).
\end  {eqnarray}
A value in a bracket denotes the estimate that gave the best
likelihood function.
The SG critical temperature coincided with the CG one.
This value disagrees with the one estimated by Fernandez et.~al,
who reported $T_{\rm SG}=T_{\rm CG}=0.120$. 
It also disagrees with the one estimated by
Viet and Kawamura, who reported $T_{\rm SG}=0.125$,
but their value $T_{\rm CG}=0.143$ is close to our estimate.
On the other hand, the value of $\nu$ is consistent with the previous 
estimates, and also consistent with the experimental results.

\begin{figure}
  \resizebox{0.40\textwidth}{!}{\includegraphics{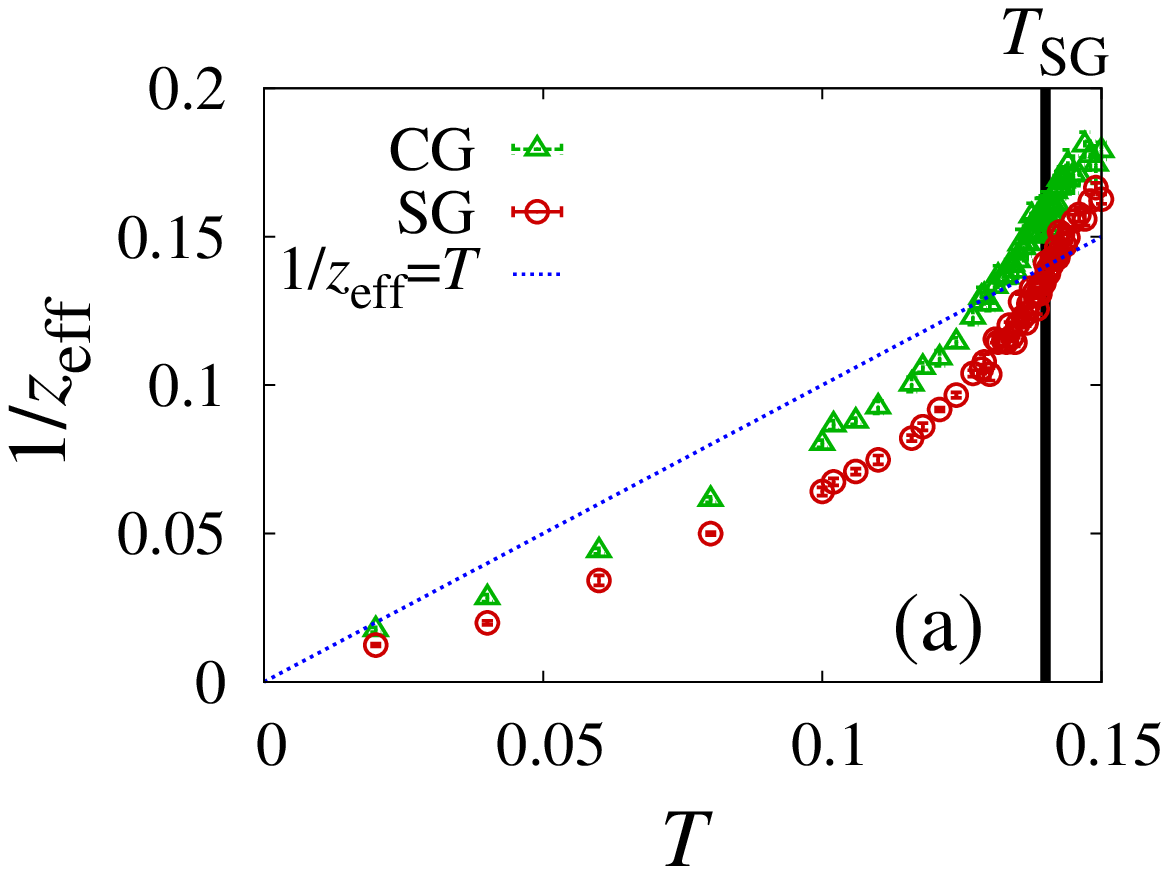}}
  \resizebox{0.40\textwidth}{!}{\includegraphics{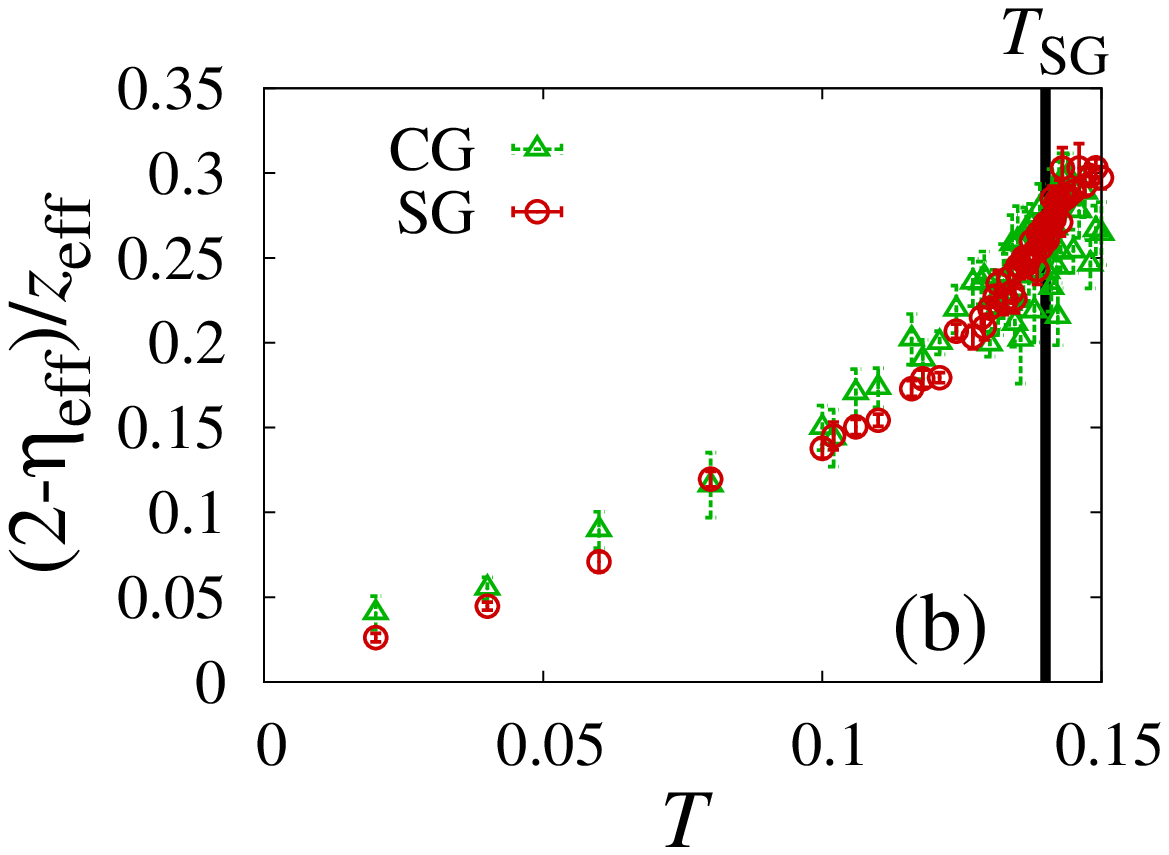}}
  \caption{
(Color online)
 Temperature dependences of effective exponents (a) $1/z_{\rm eff}$, and
 (b) $(2-\eta_{\rm eff})/z_{\rm eff}$.
}
\label{fig:z-t}
\end{figure}

Let us study a behavior of the dynamic exponent, $z$.
Since $\xi(t)\sim t^{1/z}$ in the critical region,
we can define an effective dynamic exponent, $z_{\rm eff}$, by an inverse
of a slope of Fig.~\ref{fig:namat}(b)
in the nonequilibrium process before the finite-size crossover occurred.
We estimated the value by the least-square method.
As shown in Fig.~\ref{fig:z-t}(a),
the effective dynamic exponent of SG is always larger than that of CG.
Our estimate at the transition temperature is
$z_{\rm SG}=7.3(3)$ for SG, and 
$z_{\rm CG}=6.4(2)$ for CG.
A divergence of $\xi_{\rm SG}$ is slower than that of $\xi_{\rm CG}$.
On the other hand,
a coupled exponent $z\nu$ took the same value as
$z_{\rm SG}\nu_{\rm SG}=
 z_{\rm CG}\nu_{\rm CG}=10.3$.
This agreement means that a correlation time of SG diverges with
the same speed as that of CG, because a correlation time
$\tau \sim |T-T_{\rm c}|^{-z\nu}$.
The effective dynamic exponent rapidly increased below 
the critical temperature faster than a behavior of 
$1/z \propto T$, 
which was reported\cite{katzgraber,marinari,kisker1,komori1,joh}
previously.
There is no anomaly down to the lowest temperature we simulated.
This smooth behavior is consistent with the one 
reported\cite{katzgraber} in the Ising SG model.

We also studied a temperature dependence of a coupled exponent,
$(2-\eta)/z$, which is a slope of Fig.~\ref{fig:namat}(a).
The results are plotted in Fig.~\ref{fig:z-t}(b).
This coupled exponent of SG and that of CG
behave in a same manner down to the lowest temperature.
The values at the critical temperature were 
0.266(10) for SG and
0.257(16) for CG.
This agreement means that dynamics of $\chi_{\rm SG}$ 
is equivalent to that of $\chi_{\rm CG}$, because
$\chi(t) \sim t^{(2-\eta)/z}$.

\section{Summary and Discussion}
\label{sec:discussion}

It was found 
in this paper
that a competition between the SG order and
the boundary conditions is a main origin of the difficulties in SG simulations.
As was observed in the SG profile,
a periodic boundary condition makes a strong influence to the SG spin state.
In order to succeed in equilibrium simulations of SG systems, we must
find a proper boundary condition compatible with the SG state.
However, we have not found it yet.

A periodic boundary condition produces an additional symmetry of
translating $L$ lattice spacings, which the original SG system does not have.
Spins change their state to the boundary-affected equilibrium state.
We consider that this state is quite different 
from the original SG-ordered state.
Therefore, it takes a very long time to reach the equilibrium state.
Then, the obtained data show strong finite-size effects and 
strong sample dependences.
We also found that
a size of the SG-ordered cluster is very large and hits the boundary
edge at a considerably short step:
the size reached 80 lattice spacings only at $t=10$.
The boundary-affected equilibrium state that hit the boundary
within the initial relaxation process may not include a relevant information.
Therefore, we sometimes encounter a size crossover only above which
the data should be used to study the critical phenomena.
This size crossover was first observed by
Hukushima and Campbell\cite{hukushimacampbell} who reported it in the
Ising SG model.
The correlation-length ratio changed its trend
from increasing to decreasing at a crossover size, $L=24$.

\begin{table}
\begin{center}
\begin{tabular}{l| c c c c c c}
\hline
Works & $T_{\rm SG}$ & $T_{\rm CG}$ & $\nu_{\rm SG}$& $\nu_{\rm CG}$ & $\eta_{\rm SG}$&$\eta_{\rm CG}$ \\
\hline
Present (G)& 0.140(2) & 0.140(2)& 1.4(1) & 1.6(1) & 0.04(2) & 0.40(4) \\
Ref\cite{fernandez}\hfill (G) & 0.120(6)  & 0.120(6)  & 1.5 & 1.4(1)    & -0.15(5)  & -0.75(15)    \\
Ref\cite{viet}\hfill (G) & 0.125(6)  & 0.143(3)  & - & 1.4(2) & - & 0.6(2) \\
\hline
Ref\cite{HukushimaH2}\hfill ($J$)& 0 & 0.19(1) & - & 1.3(2) & - &  0.8(2) \\
Ref\cite{totabayes}\hfill ($J$)& 0.203(1) & 0.201(1) &1.49(3) & 1.53(3) & 0.28(1) & 0.66(1) \\
\hline
Ref\cite{campbellpetit}\hfill (Ex) & & & 1.3-1.4 &    -    & 0.4-0.5 &    -   \\
\hline
\end  {tabular}
\end{center}
\caption{Comparison of present results with the previous works.
(G) stands for the Gaussian bond distribution model, 
($J$) stands for the $\pm J$ bond distribution model,
and 
(Ex) stands for experimental results.}
\label{table:tc}
\end  {table}

We confirmed that the SG transition and the CG transition
occur at the same temperature within the error bars.
A critical exponent $\gamma$ took a common value, but other critical 
exponents, $\nu$, $2-\eta$ and $z$, were different between them.
However, if we coupled exponents as $z\nu$ and $(2-\eta)/z$, 
they took common values between SG and CG.
It suggests that critical phenomena of spin glasses are better
understood by these coupled exponents.
We compared our results with the previous works in Tab.~\ref{table:tc}.
A value of $\nu_{\rm SG}$ is common between the Gaussian model and the
$\pm J$ model.
It is also consistent with a value of $\nu_{\rm CG}$.
Even if a spin anisotropy effect mixes the spin degrees of freedom and
the chirality degrees of freedom, a value of $\nu$ may not change much.
Therefore, our estimate was also consistent
with the experimental result\cite{campbellpetit}.
On the other hand, a value of $\eta $ depends much on the distribution
and on each analysis.
The SG values and the CG value also differ much.
We cannot conclude which one can explain the experimental result.

We introduced an efficient strategy avoiding the difficulties 
in SG simulations.
We consider that our strategy 
will be successfully applied to other random systems.
Here, it is essential to remove the boundary effect first.
Once the boundary effect was removed, 
the obtained data showed quite normal behaviors regardless of
whether they are nonequilibrium ones or equilibrium ones,
and
regardless of
whether the temperature is above or below the critical temperature.
A sample deviation of the SG susceptibility
vanished linearly with $1/m \to 0$, which
suggests that this value is self-averaging in this limit.
It is also noted that the error bar shrinks proportional to 
$1/({\rm computational~cost})$.

\acknowledgments
This work is supported by Grant-in-Aid for Scientific Research from
the Ministry of Education, Culture, Sports, Science and Technology, Japan
 (No. 24540413).

\end{document}